\documentclass[12pt,leqno]{article}
\usepackage[]{amsmath}
\usepackage{amsthm} 
\usepackage[]{amssymb}
\usepackage{endnotes}
\usepackage{graphicx}

\newtheorem{theorem}{Theorem}
\newtheorem{thm}{Theorem}[section]

\newtheorem{problem}[thm]{Problem}

\numberwithin{equation}{section}

\date{22 august 2019}

\begin{document}

\title{Strategic Insider Trading Equilibrium with a non-fiduciary market maker}

\author{Knut K. Aase$^{1}$,  and Bernt  \O{}ksendal$^{2}$ \\ 
\texttt{Knut.Aase@nhh.no, oksendal@math.uio.no}}


\maketitle

\footnotetext[1]{Norwegian School of Economics (NHH), Helleveien 30, N--5045 Bergen, Norway.}
\footnotetext[2]{Dept of Mathematics, University of Oslo, P.O. Box 1053 Blindern, N--0316 Oslo, Norway.}

\begin{abstract}
 The continuous-time version of  Kyle's (1985) model is studied, in which  market makers are not fiduciaries. They have some market power which they utilize to  set the price to their advantage, resulting in positive expected profits.   This has several implications for the equilibrium, the most important  being that  by setting a modest fee conditional of the order flow, the market maker is able to obtain a profit of the order of magnitude, and even better than,  a perfectly informed insider.  Our model also  indicates why speculative prices are more volatile than predicted by fundamentals. 


\medskip

\medskip

\end{abstract}

\section{Introduction}

In his seminal paper on insider trading, Albert Kyle (1985) asks several questions: How valuable is private information to an insider? How does noise trading affect the volatility of prices? What determines the liquidity of a speculative market? He provides answers to these  and other questions by modeling rigorously the trading strategy of an insider in a model of efficient price information. 

One important feature of a real securities market that remained unexplained in Kyle's analysis is the existence of a bid-ask spread.  Kyle focuses on a single auction model in which a risky asset is exchanged for a riskless asset among three kinds of traders. A single insider has access to perfect, private observation of the ex post liquidation value of the risky asset. Uninformed noise traders trade randomly.  Market makers set prices and clear the markets after observing the quantities traded by others. 

In the Kyle model the noise traders can be considered as less than fully rational, since they expect to suffer losses equal to the insiders' gains. The market makers set the prices equal to the expected value of the risky asset conditional on the order flow; they are making zero profits. The market makers cannot distinguish the trading of the insider from the trading of the noise traders, who in effect provide camouflage, which enables the insider to make profits on their expense.

The market maker in the standard model has substantial market power, yet does not exploit this to his own advantage when setting the price; the market maker is assumed to be a fiduciary acting in the best interest of market participants. 

One may ask how realistic this assumption is. In the testimony before the Financial Crisis Inquiry Commission, Goldman CEO Lloyd Blankfein laid out the Goldman Sachs perspective on the firm's role in CDO deals related to the 2008 financial crisis. From his answer it seems clear that he does not consider a market maker as a fiduciary agent:

\emph{In our market-making function, we are a principal. We represent the other side of what people want to do. We are not a fiduciary. We are not an agent. Of course, we have an obligation to fully disclose what an instrument is and to be honest in our dealings, but we are not managing somebody else's money.}

Caveat emptor seems to be Mr. Blankfein's message, and this was also the basis of Goldman's defense against the SEC suit re the Abacus transactions. The case of Goldman Sachs  is, we believe, not unique. Investment banks and other financial intermediaries  are known to accumulate large fortunes, which should be difficult, or even  impossible, if they were just disinterested auctioneers.

In this paper, we investigate the consequences of relaxing the assumption that market makers are fiduciaries. In our model, market makers are economic agents allowed to make a profit. Market makers generate profits by adding a margin to the conditional expected value of the risky asset when they are going short. Similarly they are subtracting a margin when taking long positions. Formally, the margin is a random variable, which is correlated with aggregate demand. Thus, market makers are not just adding or subtracting a fee; the size of the fee depends on trading volume.  As in the standard model, informed traders realize what market makers are up to, and take their behavior into account when deciding their own trades. Noise traders just trade, we could  allow them to have partial information as in  Aase, Bjuland and \O ksendal (2012a), but we have chosen to let these agents be uninformed. Despite of this, market makers may make unbounded profits taking advantage of noise traders, which would not make sense. To avoid this outcome a regulator is introduced.  Alternatively, the market maker may be assumed to practice restraint in order to keep markets open.

These issues  were addressed in a recent paper (Aase and Gjesdal (2017)), in the setting of a one-period model. It is  of interest to extend this analysis to several periods, which we do in this paper, where we consider  a continuous-time model.

As in the one-period model our analysis shows, perhaps surprisingly, that for only a moderate correlation with the aggregate demand,  the profits of the market maker may exceed that of the perfectly informed insider.  In the paper we illustrate the time profiles of these profits. This could serve as one explanation of why so much wealth tends to end up in the hands of financial intermediaries, a timely question that has been asked many times over after the $2008$-financial crisis.

Another implication  is that the market maker's actions leads to a more volatile price then would be the case if  dealings were fair.  This may throw some  light on the  observation made by Campbell and Schiller (1988): Stock market prices display much  more volatility than implied by  dividends alone.  The more recent approach to this problem is to consider the variations in the stochastic discount factor, see e.g., Campbell (1999). We try to see if the effects of  the actions of  market makers are substantial enough to explain parts of this problem. 

To limit the distortion of prices, a regulatory authority (the SEC)  impose an upper bound on price volatility. In our model this  limits the degree to which the market maker can perturb the price, and allows us to find an equilibrium in which the insider  maximizes profits and the market maker trades ''fees''. Even if the regulatory constraint limits the market maker's degree of price distortion, still the market maker's profit may  exceed that of the perfectly well informed insider. This happens for reasonable degrees of price distortions, a consept developed in the paper.

Our pricing functional is nonlinear, which seems like a popular topic in itself in parts of the extant literature, together with ''model uncertainty'' and similar issues. In our model the nonlinearity stems from a specific economic assumption, namely that the market maker trades fees. As we know, in neoclassical  equilibrium theory prices are linear for a variety of reasons, among others  to avoid arbitrage possibilities, which is not an issue here.

There is a rich literature on the one period model, as well as on discrete time insider trading, e.g., Holden and Subrahmanyam (1992), Admati and Pfleiderer (1988), and others, all adding insights to this class of problems.  Glosten and Milgrom (1985) present a different approach, containing similar results to Kyle. Before Kyle (1985) and Glosten and Milgrom (1985) there is also a huge literature on insider trading in which the insider acts competitively, e.g., Grossman and Stiglitz (1980).

The approach of this article is to study the continuous-time model directly, not as a limiting model of a sequence of auctions, and use the machinery of infinitely dimensional optimization, directional derivatives (or calculus of variations)  and filtering theory to solve the problem. We also consider alternatively the  stochastic maximum principle in  the setting of differential game theory, as well as the Bellman approach  in two appendices.

We are able to find the price of the risky asset, the various profits of the participants,  all in terms of  the insider's trading intensity. The latter we  show satisfies an integral equation, that can be solved by an iterative procedure. This we illustrate numerically,  by graphs of the the trading intensity,  the profits of the agents, and the other key variables developed in in the paper.

The paper is organized as follows: The model is explained in Section 2. The analysis  of the continuous time model starts in Section 3, where the mean, the variance and the covariances of the order flow $y$ is derived in Section 3.1, with preliminary expressions for the profit functions of the insider and the market maker. In Section 4 the insider's optimization problem is solved in Theorem 1, resulting in final expressions for the various profit functions, as well as the other quantities of interest.  In Section 5 we suggest how the regulator's problem may be solved, in Section 6 we introduce a measure of price informativeness in the market, and present numerical illustrations. Section 7 presents various graphs and computations, which illustrate  the key quantities in the paper. This is where we demonstrate our main conclusions. In Section 8 we provide some suggestions for further research, and Section 9 concludes. The paper also contains four appendices.

\section{The Model}
At time $T$ there is to be a public release of information that will perfectly reveal the value of an asset; cf. fair value accounting.  Trading in this asset and a risk-free asset with interest rate zero is assumed to occur continuously during the interval $[0, T]$. The information to be revealed at time $T$ is represented as a signal $\tilde{v}$,  a random variable which we interpret as the price at which the asset will trade after the release of information. This information is already possessed by a single insider at time zero. The unconditional distribution of $\tilde{v}$ is assumed to be {\em normal\/}  with mean $\mu_{\tilde{v}}$ and variance $\sigma_{\tilde{v}}^2$.

In addition to the insider, there are noise (liquidity) traders, and risk neutral market  makers. The noise traders are unable to correlate their orders to the insider's signal $\tilde{v}$. Thus the noise traders have random, price-inelastic demands. All orders are market orders and the net order flow is observed by the market maker. We denote by $z_t$ the cumulative orders of noise traders through time $t$. The process $z_t$ is assumed to be a Brownian motion with mean zero and variance rate $\sigma_t^2$, i.e.,  $dz_t  = \sigma_t d B_t$, for a standard Brownian motion $B$ defined on a probability space $(\Omega, P)$. As  Kyle (1985) and Back (1992) we assume that $B$ is independent of $\tilde{v}$.
We let $x_t$ be the cumulative orders of the informed trader, and define 
\begin{equation}\label{E: y}
y_t = x_t + z_t \qquad \text{for all $t \in [0, T]$}
\end{equation}
as the total orders accumulated by time $t$.

The market maker only observes the process $y$, so he cannot distinguish between informed and uninformed trades. Let $\mathcal{F}_t^y = \sigma(y_s; s \leq t)$ be the information filtration of this process. The risk neutral market maker, assumed to have some degree of monopoly power, sets the price $p_t$ at time $t$ as follows
\begin{equation}\label{E: price}
p_t = E[\tilde{v} + u_t|\mathcal{F}_t^y] := m_t + E[u_t|\mathcal{F}_t^y],
\end{equation}
where $m_t =  E[\tilde{v}|\mathcal{F}_t^y]$ is the "fair value", and  $u_t = k_ty_t$ for $k_t\geq 0$ a deterministic function satisfying $k_t \to 0$ as $t \to T$. We assume that $k_t = (T-t)\kappa$, where $\kappa$ is a non-negative constant set by the market maker. Clearly $E[u_t|\mathcal{F}_t^y] =k_ty_t$.
The market maker, the insider and the noise  traders all know the probability distribution of  $\tilde{v}$.

We assume that the insider's market order at time $t$ is of the form
\begin{equation}\label{E: x}
dx_t = (\tilde{v} - p_t)\beta_t dt, \qquad x_0 =0,
\end{equation}
where $\beta\geq 0$ is some deterministic function. This form of the market order follows from the discrete time formulation of the problem, assuming the insider maximizes  profits, in which case (\ref{E: x}) follows from the first order condition; $x_t$ does not depend on $p_t$ since $x_t$ is submitted before $p_t$ is set by the market makers. 

Assumption  (\ref{E: x}) is consistent with Kyle (1985).\footnote{The finite variation property of $x$ is assumed by Kyle (1985), and an equilibrium where this is the case is found  by Back (1992).} The function $\beta_t$ is called the  \emph{trading intensity} on the information advantage $(v - p_t)$ of the insider. 

The  basic assumptions behind this result is (i) profit maximization by the insider,  where it is shown in Aase and Gjesdal (2016) that this result still holds when the market maker sets the price as we have assumed in (\ref{E: price})
 above, and (ii) the insider does not condition the quantity he trades on price. Here the insider chooses quantities ("market orders") instead of demand functions ("limit orders").

Assumption (\ref{E: price}) is our deviation from the standard model.\footnote{ An alternative would be to assume  that the  market maker is risk averse, which would lead to a different model. } Below we explain why this price setting leads to a positive expected profit for the market maker.

The stochastic differential equation for the total order $y_t$ is 
\begin{equation}\label{E: yt}
dy_t = (\tilde{v} - m_t)\beta_t dt - k_t\beta_ty_t\,dt +\sigma_t\, dB_t.
\end{equation}
Aside from the first mean zero 'innovation' term, the equation shows that $y_t$ has the structure of  a (generalized) mean reverting Ornstein-Uhlenbeck process, oscillating  around this mean zero term.

Let us denote by $S_t(\beta) = E\{(\tilde{v} - p_t)^2\}$ and by $\gamma_t(\beta) = E\{(\tilde{v} - m_t)^2\}$.    Usually the assumption is made that $\lim_{t\to T^-}p_t = p_T = \tilde{v}$ a.s.  This assumption seems natural, ensuring that all information available has been incorporated in the price at the time $T$ of the public release of the information, at which time a price spread cannot be sustained. 

In Aase et. al. (2012a)  $m_t = p_t$ for all $t \in [0,T]$, and it was there demonstrated that $p_t \to \tilde{v}$ as $t \to T^-$, and $S_t(\beta) \to 0$ as $t \to T^-$ as a consequence of the insider following his optimal trading strategy. Here we find it natural to simply assume this, as was done in e.g., Kyle (1985),   so that $p_t - m_t \to 0$ as $t \to T^-$, and both converge to $\tilde{v}$, since $k_t \to 0$ by assumption.

Denote the insider's wealth by $w$ and the investment in the risk-free asset by $b$. The budget constraint of the insider can best be understood by considering a discrete time setting, of which the continuous-time model is the limit (in an appropriate sense). At time t the agent submits a market order $x_t - x_{t-1}$ and the price changes from $p_{t-1}$ to $p_t$. The order is executed at price $p_t$, in other words, $x_t - x_{t-1}$ is submitted \emph{before} $p_t$ is set by the market makers. The investment in the risk-free asset changes by $b_t - b_{t-1} = -p_t(x_t - x_{t-1})$, i.e., buying stocks leads to reduced cash with exactly the same amount. Thus, the  associated change in wealth is 
\begin{equation}\label{E: partial sum}
b_t - b_{t-1} +  x_t p_t - x_{t-1}p_{t-1} = x_{t-1}(p_t - p_{t-1}).
\end{equation}
In other words, the  accounting identity for the wealth dynamics is of the same  type as in the standard price-taking model, except for one important difference; while, in the rational expectations model,  the number of stocks in the risky asset at time $t$ depends only on the information available at this time, so that both the processes $x$ and $p$ are adapted processes with respect to the same filtration, here the order $x$ depends on information available only at time $T$ for the market makers (and the noise traders). 

As a consequence  of (\ref{E: partial sum})  we obtain the dynamic equation for the insider' wealth $w_t^{I}$ as follows
\begin{equation}
w_t^{I} = w_0^{I} + \int_0^t x_{s}dp_s
\end{equation}

This is not well-defined as a stochastic integral in the traditional interpretation, since $p_t$ is $\mathcal{F}_t^y$-adapted, and  $x_t$ is not. Thus it needs further explanation.
However, since we assume that the strategy of the insider has the form (2.3)
for some deterministic continuous function $\beta_t >0$, then a natural interpretation of (2.6)
is obtained by using \emph{integration by parts}, as follows:

\begin{align}
w_t^{I}&=w_0^{I}+ x_t p_t -\int_0^t p_s dx_s \nonumber\\
&=w_0^{I} + p_t\int_0^t (\tilde v -p_s)\beta_s ds -\int_0^t p_s (\tilde v -p_s)\beta_sds
\nonumber\\
&=w_0^{I} + \int_0^t (\tilde v -p_s)^2\beta_s ds
-\int_0^t (\tilde v -p_t)(\tilde v -p_s)\beta_sds.
\label{eq1.6}
\end{align}

Alternatively, one might obtain \eqref{eq1.6} by interpreting the stochastic
integral in (2.6) as a \emph{forward integral}. See Russo and Vallois (1993), Russo and Vallois (1995, 2000) for
definitions and properties and Biagini and $\O$ksendal (2005) for applications of forward integrals to finance.

Similarly we can find the market maker's profit from his price setting operations: His wealth $w^M$ from these operations is
\[
w^M = w^M_0 + (p_0 - p_1)y_0 + (p_1 - p_2)y_1 + \cdots
\]
When the total initial order $y_0 > 0$,  the market maker has to sell to clear the market and accordingly  sets the price $p_0$ a bit higher than he would have done if he were a fiduciary. Similarly, if $y_0 < 0$ she must buy to clear the market, so he  sets the price $p_0$ a bit lower than he would if he  sets the price fairly. Continuing this practice in every period, he will end up with a positive expected profit, simply because the profit he would have obtained by being fair has zero expectation\footnote{One may think of trade as "synthetic" in that only money changes hands, based on dynamics of the underlying stock.}.

Consider the situation where the total initial order $y_0 > 0$.  Because of the mean reverting nature of $y$ towards zero, it is more likely that $y_1 < y_0$ than the other way around. By the price setting mechanism used by the market maker, it is more likely that $p_1 < p_0$ than the opposite,  in which case the market maker's profit is positive. A similar reasoning holds when $y_0 < 0$, in which case the market maker buys from the other participants at time zero, and sells the stock in the market at time one at the price $p_1$ he sets then, based on $y_1 - y_0$.  Thus, in expectation the market maker's profit is positive.

Notice that the market maker takes some 'overnight' risk, in that, when he must sell to the other participants at time $t$, he sets the price $p_t$ which he sells for, and the next day he sets the price $p_{t + 1}$, based on the order $y_{t + 1} - y_t$, at which he buys in the market the stock that he 'delivered' the day before. By the price setting mechanism,  more likely than the other way he profits from this operation.  If he were a fiduciary, he would go even in 'the long run'. Here as a non-fiduciary, in expectation his profit is positive. 

By going to the continuous time limit, his wealth at time $t$ is
\begin{equation}\label{E: mmp}
w_t^M = w_0^M - \int_0^t \,y_s dp_s = w_0^M -  p_ty_t + \int_0^t \,p_s dy_s + [p,y]_t,
\end{equation}
where $[p,y]_t$ is the quadratic covariance process of $p$ and $y$.
Unlike the corresponding expression for the insider, this integral is well-defined in the traditional interpretation, since $p_t$ is $\mathcal{F}_t^y$-adapted, and so is of course  $y_t$. 

Finally, the noise traders' profit is 
\begin{equation}\label{E: ntp}
w_t^N = w_0^N + \int_0^t z_s dp_s = w_0^N + z_tp_t - \int_0^tp_sdz_z -  [p,z]_t.
\end{equation}
The stochastic integral $\int_0^t z_s dp_s $ is well-defined in the traditional meaning since $z_t$ is $\mathcal{F}_t^B$-adapted, $p_t$ is $ \mathcal{F}^y_t$-adapted and  $\mathcal{F}^y_t\supset \mathcal{F}_t^B$, and hence, by integration by parts, so is the latter stochastic integral in (\ref{E: ntp}). 

Since $y_t = x _t+ z_t$ and $x$ is of bounded variation,  $[p,y]_t =  [p,z]_t$ for all $t$. Since this is a pure exchange economy, it follows that the sum of the profits is zero with probability one, or, $w_t^{I} + w_t^{M}+ w_t^{N} = w_0^{I} + w_0^{M} + w_0^{N}$ a.s.

\section{Some basic analysis.}
Returning to the stochastic process for the total order at time $t$, $y_t$, its representation is given by (\ref{E: yt}), which we repeat here
\[
dy_t = \big(\tilde{v} - E(\tilde{v}|\mathcal{F}_t)\big)\beta_t dt - k_t\beta_ty_t\,dt +\sigma_t\, dB_t.
\]
This is a Gaussian process  consisting of an Ornstein-Uhlenbeck type process, with  a  normally distributed  ''innovation'' term added to its drift term, the first term on the right-hand side in the above stochastic differential equation.
 
In order  to analyse this model for the total order, we start by rewriting this equation as follows:
\begin{align*}\label{E: yt2}
dy_t+y_tk_tdt=(\tilde{v}-m_t)\beta_tdt + \sigma_t dB_t.
\end{align*}
If we define
\begin{equation}\label{E: tilde y}
\tilde{y}_t:=y_t \exp(\int_0^t k_s \beta_s ds)
\end{equation}
and use Ito's lemma, we obtain the following
\begin{align}\label{eq1.1}
d\tilde{y}_t=(\tilde{v}-m_t)\beta_t \exp(\int_0^tk_s\beta_s ds) dt + \sigma_t \exp(\int_0^tk_s\beta_s ds)dB_t.
\end{align}
Clearly $\mathcal{F}^{(y)}=\mathcal{F}^{(\tilde{y})}$ and hence
\begin{equation}\label{E: mt}
m_t= E[\tilde{v} | \mathcal{F}^{(y)}] = E[\tilde{v} | \mathcal{F}^{(\tilde{y})}].
\end{equation}
Therefore we may regard \eqref{eq1.1} as the innovation process of an "observation process" $\hat{y}_t$ defined by
\begin{align}\label{eq1.2}
d\hat{y}_t=\tilde{v}\beta_t\exp(\int_0^t k_s\beta_s ds)dt + \sigma_t \exp(\int_0^t k_s\beta_s ds) dB_t; \quad \hat{y}_0=0.
\end{align}
For this to hold, we need to verify that
\begin{equation} \label{eq1.3}
\mathcal{F}_t^{(\tilde{y})}=\mathcal{F}_t^{(\hat{y})}.
\end{equation}
Suppose \eqref{eq1.3} is proved. Then
$$ m_t= E[\tilde{v}| \mathcal{F}_t^{(\hat{y})}]$$
is the filtered estimate of $v$ given the observations $\tilde{y}_s; s \leq t$.

By Theorem 12.1 in [18] or Theorem 6.2.8 in [19], the filter $m_t$ is given by the SDE
\begin{align}\label{eq1.4}
dm_t&=\frac{\gamma_t \beta_t \exp(\int_0^tk_s\beta_s ds)}{\sigma_t^2 \exp(2 \int_0^t k_s\beta_s ds)} \Big[d\hat{y}_t - \beta_t \exp(\int_0^tk_s\beta_s ds)m_t dt \Big ];\quad t \geq 0; \\
&m_0=E[\tilde{v}] \nonumber,
\end{align}
where $S_t=S_t^{(\beta)} = \gamma_t(\beta) + k_t^2V(t)$, where $V(t) = E(y_t^2)$, and $\gamma_t(\beta)$ solves the Riccatti equation
\begin{align}\label{eq1.5}
d\gamma_t&= - \frac{\beta_t^2 \gamma_t^2}{\sigma_t^2}; \quad t \geq 0  \\
\gamma_0&=E[(\tilde{v}-E[\tilde{v}])^2]  \nonumber.
\end{align}
Thus we have a controlled state process
$$(\hat{y}_t, m_t, \gamma_t)$$
given by the equations \eqref{eq1.2},\eqref{eq1.4} and \eqref{eq1.5}.

\vskip 0.3cm

Rewriting the system in terms of $(y_t,m_t,\gamma_t)$ we obtain the following set of equations
\begin{align}
&dy_t=(\tilde{v}-m_t-k_ty_t)\beta_t dt +\sigma_t dB_t;\quad y_0=0 \\
&dm_t=\frac{\gamma_t\beta_t}{\sigma_t^2}[(\tilde{v}-m_t)\beta_t dt + \sigma_t dB_t];\quad m_0 = p_0 = E[\tilde{v}]\\
&d\gamma_t=- \frac{\beta_t^2 \gamma_t^2}{\sigma_t^2}; \quad \gamma_0=E[(\tilde{v}-E[\tilde{v}])^2].
\end{align}
The expected profits are
\begin{align}
&J^{M}(k,\beta):=w_0^M + E(\int_0^T\, k_t y_t\,( k_ty_t  + m_t - \tilde{v})\beta_t dt - \int_0^Ty_t^2\, dk_t)\\
&J^{I}(k,\beta):= w_0^{I}+ \int_0^T E[(\tilde v -m_s - k_sy_s)^2]\beta_s ds.
\end{align}
for the market maker and for the insider, respectively.

Let us now return to the problems of the previous section and calculate the profits of various participants in this economy.  Towards this end we first need  expressions for the mean, the  variance and the covariances of the market order process $y$.

\subsection{The variance and covariances of  the process $y$.}
We start with the variance. Based on the expression in (\ref{E: mt}), we proceed as follows. From equation (\ref{E: tilde y})   we have by It\^{o}'s lemma
\[
d(\tilde{y}_t)^2 = 2\tilde{y}_td\tilde{y}_t + \frac{1}{2}2(d\tilde{y}_t)^2 =
\]
\[
2y_t^2\big[(\tilde{v} - m_t)\beta_t \text{exp}\big(\int_0^tk_s\beta_sds\big)dt + \sigma_t\,\text{exp}\big(\int_0^tk_s\beta_sds\big)dB_t\big] +
\]
\[
\sigma_t^2\,\text{exp}\big(2\int_0^tk_s\beta_s\big)dt.
\]
From this we deduce that
\[
E\big[\tilde{y}_t^2\big] = E\Big[\int_0^t2y_s^2\Big[(\tilde{v} - m_s)\beta_s \text{exp}\big(\int_0^sk_u\beta_udu\big)ds
\]
\[
 + \sigma_s\,\text{exp}\big(\int_0^sk_u\beta_udu\big)dB_s\big] +\sigma_s^2\,\text{exp}\big(2\int_0^s k_u\beta_udu\big)ds\Big] =
\]
\[
\int_0^t\sigma_s^2\,\text{exp}\big(2\int_0^sk_u\beta_udu)ds.
\]
Observe that
\[
E\big((\tilde{v} - m_t)y_t^2\big) = E(\tilde{v}y_t^2) -  E\big(E(\tilde{v}|\mathcal{F}_t^y)y_t^2\big) = 0
\]
since $E\big(E(\tilde{v}|\mathcal{F}_t^y)y_t^2\big) = E\big(E(y_t^2\tilde{v}|\mathcal{F}_t^y)\big) = E(\tilde{v}y_t^2)$.
Hence
\[
\text{exp}\big(2\int_0^tk_u\beta_udu\big)E\big[y_t^2\big] = \int_0^t\sigma_s^2\, \text{exp}(2\int_0^sk_u\beta_udu\big)ds
\]
or
\begin{equation}\label{E: eysquare}
E\big[y_t^2\big] = e^{-2\int_0^tk_r\beta_rdr}\int_0^t\sigma_u^2\, e^{2\int_0^u k_r\beta_rdr}du.
\end{equation}
This expression will be useful below. We use the notation $V(t) := E(y_t^2)$ for all $t \in [0,T]$. 
\footnote{The theory leading to the result in (\ref{E: eysquare})  may be linked to a deeper result in filtering theory. For details, see Appendix 4.}

Moving to the covariances $E(y_ty_s)$ for any $s > t$, we proceed as follows. Here we use the notation 
\[
e(t) := e^{\int_0^tk_r\beta_rdr}.
\]
For $s > t$,
\[
d_s(\tilde{y}_s\tilde{y}_t) = (\tilde{v} - m_s)\beta_s e(s) \tilde{y}_tds + \sigma_s e(s)\tilde{y}_t dB_s.
\]
Integrating this from $t$ to $s$, we get
\[
E[(\tilde{y}_s - \tilde{y}_t)\tilde{y}_t] = E[\int_t^s(\tilde{v} - m_r)\beta_re(r)\tilde{y}_tdr + \int_t^s\sigma_re(r)\tilde{y}_tdB_r] =
\]
\[
\int_t^sE[(\tilde{v} - m_r)\tilde{y}_t]\beta_re(r)dr + 0 = 0,
\]
since 
\[
E[(\tilde{v} - m_r)\tilde{y}_t] = E[E[(\tilde{v} - m_r)\tilde{y}_t | \mathcal{F}_t^y] = E[\tilde{y}_tE[\tilde{v} - m_r|\mathcal{F}_t^y]] = 0.
\]
The latter equality follows from $E[(\tilde{v} - m_r)|\mathcal{F}_t^y] = E[E[(\tilde{v} - m_r)|\mathcal{F}_r^y|\mathcal{F}_t^y]] = 0$, since the inner conditional expectation is zero.
We obtain  for $s > t$ 
\[
E[\tilde{y}_s\tilde{y}_t] = E[(\tilde{y}_s - \tilde{y}_t)\tilde{y}_t] + E[\tilde{y}_t^2] = E[\tilde{y}_t^2].
\]
Using the definition of $\tilde{y}$, we have that 
\[
E[y_sy_t]  = E[y_t^2]e^{-\int_t^sk_r\beta_rdr}
\]
Combining this with our above result (\ref{E: eysquare}), we conclude that
\begin{equation}\label{covy}
E[y_sy_t]  = e^{-(\int_0^s k_r\beta_rdr + \int_0^tkr\beta_rdr)}\int_0^t\sigma_u^2e^{2\int_0^uk_r\beta_rdr}du, \quad \text{for} \,\,s > t.
\end{equation}
For $s = t$ we obtain the result in equation (\ref{E: eysquare}).

Figure 1 illustrates a graph of the covariance function $C(s,t;\kappa):= E[y_sy_t]$ when $\kappa = 0.045$ for $s, t \in[0,10]$.  

\vbox{\centerline{\includegraphics[width=45mm]{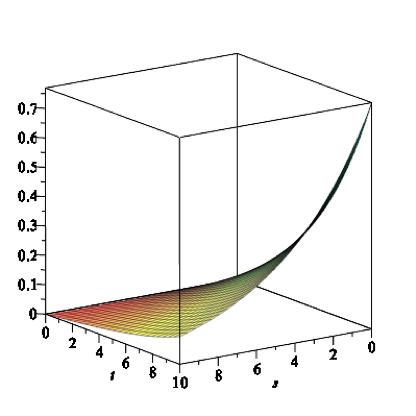}}
\centerline{Fig. 1: The covariance function $C(s,t)$ of $y$ when $\kappa= 0.045.$}}

The base case parameters  are $\sigma_t = \sigma = 0.20$, a constant for all $t \in [0, T]$. Also $\gamma_0 = \sigma_{\tilde{v}}^2$, where $\sigma_{\tilde{v}} = 0.30$, and we have chosen $T = 10$. (Here we have anticipated a bit, and used the optimal value of the trading intensity $\beta_t$ of the insider appearing in Section 4 below.)

\subsubsection{ The mean of y}
We will also need the mean $E(y_t)$ of the process $y$ for any $t$. Starting with the equation
\[
y_t = y_0 + \int_0^t(\tilde{v} - E(\tilde{v}|\mathcal{F}_s)\beta_sds - \int_0^t\,k_s\beta_sy_sds + \int_0^t\sigma_sdB_s,
\]
and letting $E(y_t) := \bar{y}_t$, where $\bar{y}_0 = y_0$, by taking expectation in the above equation we obtain
\[
 \bar{y}_t = y_0 - \int_0^tk_s\beta_s\bar{y}_sds
 \]
 or 
 \[
 \frac{d}{dt}\bar{y}_t = - \beta_tk_t\bar{y}_t
 \]
which is an ordinary, linear differential equation in $\bar{y}_t$, with initial condition $\bar{y}_0 = y_0$, the unique solution of which is
\[
E(y_t) = y_0e^{-\int_0^tk_s\beta_sds}.
\]
In our model $y_0 = 0$, which implies that $E(y_t) = 0$ for all $t \in [0,T]$. Thus $E(p_t) = E(m_t) + k_tE(y_t)  = E(m_t) = E(\tilde{v}) = \mu_{\tilde{v}}$, so the price $p_t$ has the correct expectation at all times.

\subsection{The profit of the insider}

Returning to the insider,  from equation (\ref{eq1.6}) giving the wealth $w_t$ of the insider at any time $t$, since
\[
\int_0^T E[(\tilde v -p_T)(\tilde v - p_t)]\beta_tdt = 0
\]
by our assumption that $p_t \to p_T = \tilde{v}$, his task is to find the trading intensity $\beta_t$ which maximizes the expected terminal wealth
\begin{equation}
E[w_T^I]= w_0^I + \int_0^T E[(\tilde v -m_t - k_ty_t)^2]\beta_t dt:= J^I(k,\beta).
\label{eq1.7}
\end{equation}
Later, when we consider the net profit at any time $t \in [0,T]$, we will use the notation $p_I(t, \kappa)$ for the insider's net profit by time $t$, so that $J^I(k, \beta ) - w_0^I:= p_I(T, \kappa))$ with this notation. Similarly for the market maker.

The dilemma for the insider is that an increased trading intensity at some time $t$
will reveal more information about the value of $\tilde{v}$ to the market makers and hence
induce a price $p_t$ closer to $\tilde{v}$, which in turn implies a reduced insider information
advantage. On the other hand she has to trade in order to make any profit at all.

First observe that
\[
E\big((\tilde{v} - m_t)y_t\big) = E(\tilde{v}y_t) -  E\big(E(\tilde{v}|\mathcal{F}_t^y)y_t\big) = 0
\]
since $E\big(E(\tilde{v}|\mathcal{F}_t^y)y_t\big) = E\big(E(y_t\tilde{v}|\mathcal{F}_t^y)\big) = E(\tilde{v}y_t)$, a result similar to the one obtained above, with $y_t$ instead of $y_t^2$.

By the definition of $\gamma_t(\beta) = E(\tilde{v} - m_t)^2$,  we then obtain the following
\begin{equation}\label{E: JI}
J^I(k, \beta) = w_0^I + \int_0^T\beta_t(\gamma_t(\beta) + k_t^2V_t)dt
\end{equation}
since the cross term vanishes by by the above observation. Using the expression for $V(t):= E( y_t^2)$ given in (\ref{E: eysquare}), we obtain the following 
\begin{equation}\label{E: JI2}
J^I(k, \beta) = w_0^I + \int_0^T\beta_t\big(\gamma_t(\beta)  + k_t^2 e^{-2\int_0^tk_s\beta_sds}\int_0^t\sigma_s^2 e^{2\int_0^sk_r\beta_rdr}ds\big)dt.
\end{equation}

The insider will now maximize this expression in the trading intensity process $\beta$, for a given price perturbation process $k$ by the market maker. 

Before we address this problem, we  want to find the profit of the insider \emph{at any time} $t \in [0,T]$, which will allow us to observe the relative time performance of the two profit functions  of interest.

Towards this end, let us go back to the expression for the insider's profit at time $t$ given in (\ref{eq1.6}). Taking expectation in this equation we obtain
\[
E(w_t^I) = w_0^I + \int_0^t E(\tilde{v} - p_s)^2 \, \beta_s\,ds - \int_0^tE(\tilde{v} - p_t)(\tilde{v} - p_s)\, \beta_s\,ds =
\]
\[
w_0^I + \int_0^t (\gamma_s(\beta) + k_t^2V(s)) \, \beta_s\,ds - \int_0^tE(\tilde{v} - m_t - k_ty_t)(\tilde{v} - m_s - k_sy_s)\, \beta_s\,ds,
\]
where the second term follows from (\ref{E: JI2}). Consider the  last term. The integrand can be written 
\begin{multline}\label{E: iterated exp}
E(\tilde{v} - m_t - k_ty_t)(\tilde{v} - m_s - k_sy_s) = E(\tilde{v} - m_t)((\tilde{v} - m_s)  - \\
k_sE((\tilde{v} - m_t )y_s) - k_tE((\tilde{v} - m_s )y_t) + k_tk_sE(y_ty_s).
\end{multline}
The second expectation on the right-hand side is
\begin{multline*}
E((\tilde{v} - m_t )y_s) = E[E[(\tilde{v}  - m_t)y_s|\mathcal{F}_s^y]] =\\
E[y_sE(\tilde{v} - m_t|\mathcal{F}_s^y)] = E[y_sE[E(\tilde{v} - m_t|\mathcal{F}_t^y)]|\mathcal{F}_s^y] = 0
\end{multline*}
by standard iterated expectations, since $E(\tilde{v} - m_t|\mathcal{F}_t^y) = 0$, as shown before. 

Notice that $s \leq t$ in these computations. The third expectation on the right-hand side  of (\ref{E: iterated exp}) is
\begin{multline*}
E((\tilde{v} - m_s )y_t) = E[E[(\tilde{v}  - m_s)y_t|\mathcal{F}_s^y]]\\
 = E[E[E(y_t(\tilde{v} - m_t)|\mathcal{F}_s^y]|\mathcal{F}_t^y] 
= E[y_tE[E(\tilde{v} - m_s|\mathcal{F}_s^y)]|\mathcal{F}_t^y] = 0,
\end{multline*}
where the second equality above follows from a not so standard, but rather obvious, iterated expectation result (see e.g., Tucker (1967), Th 6, Ch 7),
and again,  because $E(\tilde{v} - m_s|\mathcal{F}_s^y) = 0$,  the result follows. 

It remains to compute the first expectation on the right-hand side  of (\ref{E: iterated exp}). It follows from Theorem 3.1 in Aase  and  \O{}ksendal (2018) that
\[
E(\tilde{v} - m_t)((\tilde{v} - m_s) = \gamma_t(\beta).
\]
The last term in (\ref{E: iterated exp}), the covariance, we have already computed in Section 3.1. Since here $t \geq s$, we rewrite this formula accordingly, namely as
\begin{equation}\label{covy2}
E[y_ty_s]  = e^{-(\int_0^t k_r\beta_rdr + \int_0^skr\beta_rdr)}\int_0^s\sigma_u^2e^{2\int_0^uk_r\beta_rdr}du, \quad \text{for} \,\,t \geq s.
\end{equation}

This means that the insider's profit at any time $t$ in $[0, T]$ is given by
\[
E(w_t^I) = w_0^I + \int_0^t(\gamma_s(\beta) + k_s^2V_s)\beta_s ds - \gamma_t(\beta)\int_0^t\beta_sds  - k_t\int_0^t k_sE(y_ty_s)ds.
\]
Observe that as $t \to T$, this profit converges to the expression in (\ref{E: JI}), since both $\gamma_t(\beta) \to 0$ and $k_t \to 0$ then.
By use of (\ref{covy2}) the insider's profit can be written
\begin{multline}\label{E: dynprofit}
E(w_t^I) = w_0^I + \int_0^t(\gamma_s(\beta) + k_s^2V_s)\beta_s ds - \gamma_t(\beta)\int_0^t\beta_sds\\
  - k_te^{-\int_0^t k_r\beta_rdr }\int_0^t\big(e^{-\int_0^s k_r\beta_rdr}\int_0^s\sigma_u^2e^{2\int_0^uk_r\beta_rdr}du\big)k_sds.
  \end{multline}
  
  The problem of finding the optimal value of the insider's trading intensity $\beta_t$, and the corresponding expression for the profit fundtion is relegated to Section 4 below.

\subsection{The profit of the market maker}
The market maker's expected profit is:
\[
J^M(k, \beta) := E(w_T^M) = w_0^M - E\big(\int_0^T\, y_t\,dp_t\big)= 
\]
\[
w_0^M - E\big(\int_0^T\, k_t y_t\,dy_t + \int_0^Ty_t^2\, dk_t\big) =
\]
\[
w_0^M - E\big(\int_0^T\, k_t y_t\,(\tilde{v} - m_t -k_ty_t)\beta_t dt + \int_0^Ty_t^2\, dk_t\big) =
\]
\[
w_0^M + \int_0^T\, k_t^2 (Ey_t^2)\beta_t dt + \kappa \int_0^TEy_t^2\, dt.
\]
The third equality follows since $m$ is a martingale, the fourth since $B_t$ is a $F_t^y$-martingale, and the last equality follows since $y_t$ is orthogonal to $(\tilde{v} - m_t)$, and the Fubini theorem.
 Thus we have that this profit can be written
 \begin{equation}\label{E: M1}
J^M(k, \beta) =  w_0^M + \int_0^T\big( k_s^2 V(s)\beta_s+ \kappa V(s)\big)ds.
\end{equation}
Notice that the profit of the market maker at any time $t \in [0,T]$ is simply 
\begin{equation}\label{E: dynmmmprofit}
E(w_t^M) = w_0^M + \int_0^t\big( k_s^2 V(s)\beta_s+ \kappa V(s)\big)ds.
\end{equation}

Using the expression (\ref{E: eysquare}) for $V_s = E(y_s^2)$, we obtain the following expression for this profit:
\begin{multline}\label{E: M}
J^M(k, \beta) =  w_0^M + \int_0^T \Big((k_s^2 e^{-2\int_0^sk_r\beta_rdr}\int_0^s\sigma_u^2 e^{2\int_0^uk_r\beta_rdr}du)\beta_s \\+ \kappa(e^{-2\int_0^sk_e\beta_rdr}\int_0^s\sigma_u^2 e^{2\int_0^uk_r\beta_rdr}du)\Big)ds.
\end{multline}

Consider the latter profit. The last term on the right-hand side increases without bounds as $k_t = (T - t)\kappa$ increases without bound for any given $t$, i.e., as the constant $\kappa\to \infty$.  Surely $k_t$ goes to zero as $t$ goes to $T$, but the constant $\kappa$ can in principle be  set arbitrarily large by the market maker, since she simply decides the value of this constant once and for all. Also we know that $\beta_t$ decreases with $\kappa$, but this effect  more or less cancels out since the two exponentials where $\beta$ enters are of different signs.

Likewise, the second term on the right-hand side, $\int_0^T\, k_t^2 (Ey_t^2)\beta_t dt$,  also possesses  this property, despite the fact that here $\beta$ enters linearly (in addition to its exponential dependence). 

This is illustrated numerically  in Figure 2. The base case  parameters are the same as in Figure 1, where the horizon is $T=10$.  (Again we have anticipated a bit, and used the optimal values of the function $\beta_t$ appearing is Section 4 below.)

 Using the notation for the net profit of the market maker
\begin{multline*}\label{E: M}
p_M(t, \kappa): =   \int_0^t \Big((k_s^2 e^{-2\int_0^sk_r\beta_rdr}\int_0^s\sigma_u^2 e^{2\int_0^uk_r\beta_rdr}du)\beta_s \\+ \kappa(e^{-2\int_0^sk_e\beta_rdr}\int_0^s\sigma_u^2 e^{2\int_0^uk_r\beta_rdr}du)\Big)ds.
\end{multline*}
at the intermediate time $t \leq T$, the upper graph is the net, terminal profit $p_M(T, \kappa)$  as a function of $\kappa$, while the  the lower graph shows the net profit $p_M(t, \kappa)$ accumulated at the intermediate time $t = 2$ as a function of $\kappa$. \\

\vbox{\centerline{\includegraphics[width=45mm]{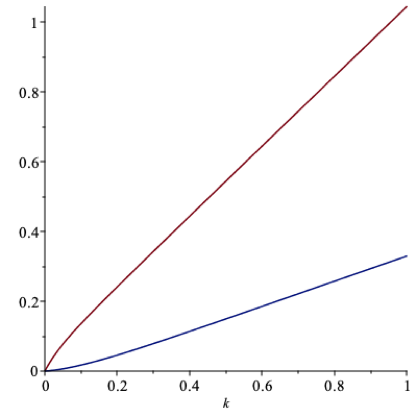}}
\centerline{Fig. 2: The profit functions of the market maker  as a function of $\kappa$.}}

 As a result, this model displays similar properties to the one-period model, and a regulator is therefore introduced to limit the price perturbation caused by the market maker trading fees.

This set-up does not become a game between the insider and the market maker in the usual meaning of game theory, in that only the insider  maximizes an objective, while the market maker trades fees that depend on the stochastic order flow, i.e., she sets the price to the best of her knowledge, and then adds the fee conditional on observing the order flow. In some sense, the market maker is not strategic  in the ordinary interpretation of this term.
 
As the market maker obtains more information from the order flow, she  lets this information be reflected in the price $p_t$. The introduction of trading fees reduces  the informational contents of the true value of the asset in the price.  The market maker may be assumed to set $\kappa$ to the maximum value allowed by the regulator, or alternatively,  by her own conscience, supposing  she practices restraint in order to keep the markets open, whichever gives the smallest  value of $\kappa$.   It is in the interests of the market maker that the market does not break down, in which case she does not make any  profits at all, and may also face legal issues. It is, after all, the market maker's task to operate such that the markets function.

The problem of relating the parameter $\kappa$ to observables in the market is treated in Section 5 below.

Since this is a pure exchange economy, the profit of the noise traders is given by
\[
J^N(k, \beta) = w_0^I + w_0^M + w_0^N -J^I(k, \beta) - J^M(k, \beta).
\]
They will loose in this market.

\section{The insider's problem}
We now address the optimization problem of the insider. In our framework he is to determine the trading intensity $\beta_t$ by which he trades  at each time $t \in [0,T]$. We assume he determines this intensity such that his profit $J^I(k,\beta)$ is maximized, taking $k$ as given. Vi have that 
\[
dp_t = dm_t + d(k_ty_t) = dm_t  -\kappa y_t dt + k_t dy_t,
\]
since the function $k_t$ is of bounded variation. From filtering theory (see e.g., Kalman (1960), Davis (1977-84), Kallianpur (1980) or \O ksendal (2003), Ch 6)   we know that the corresponding conditional expected value $m_t = E(\tilde{v}|\mathcal{F}_t^y)$ is given by 
\[
dm_t = \frac{\beta_t\gamma_t(\beta)}{\sigma_t^2}dy_t.
\]
Furthermore the square error function $\gamma_t(\beta) = E(\tilde{v} - m_t)^2$ satisfies the Ricatti equation
\[
\frac{d}{dt}\gamma_t(\beta) = -\frac{\beta_t^2}{\sigma_t^2}\gamma_t^2(\beta),
\]
which has the solution
\begin{equation}\label{E: gamma}
\gamma_t(\beta) = \frac{\sigma_{\tilde{v}}^2}{1 + \sigma_{\tilde{v}}^2\int_0^t\tilde{\beta}_s^2ds},
\end{equation}
where $\tilde{\beta}_t := \frac{\beta_t}{\sigma_t}$. Here $\gamma_0 = E(\tilde{v} - E\tilde{v})^2 = \sigma_{\tilde{v}}^2$. Accordingly, the insider's problem is to solve the following  
\begin{equation}\label{P:1}
\text{sup}_{\beta}\int_0^T\Big(\frac{\sigma_{\tilde{v}}^2\beta_t}{1 + \sigma_{\tilde{v}}^2\int_0^t\frac{\beta_s^2}{\sigma_s^2}ds} + \beta_tk_t^2 e^{-2\int_0^tk_s\beta_sds}\int_0^t\sigma_s^2 e^{2\int_0^sk_r\beta_rdr}ds\Big)dt.
\end{equation}
We find it natural to use directional derivatives, or equivalently, variational calculus to solve this problem. Based on this we have the following:

\begin{theorem}\label{T: 1}
The optimal trading intensity $\beta_t$ of the insider satisfies the following integral equation
\begin{equation}\label{E: opt beta}
\beta_t = \frac{\sigma_t^2}{2\int_t^T\gamma_s(\beta)^2\beta_s ds}\Big(\gamma_t(\beta) - V(t)\big(k_t^2+
2k_t\int_t^T\beta_sk_s^2e^{-2\int_t^sk_r\beta_rdr}ds\big)\Big),
\end{equation}
where $V(t)$ is the variance process of the order flow $y_t$.
\end{theorem}
\underline{Proof:} The proof can be found in Appendix 1.\footnote{The problem may alternatively be formulated in terms of a stochastic differential game between the insider and the market market maker, in which case we make use of the stochastic maximum principle. This leaves  
three adjoint variables (co-variables with shadow price interpretations) to be determined. Alternatively we can formulate the problem as a dynamic programming problem and use the Bellman approach. In this case this leaves us with the indirect profit  function to be determined. We indicate these two formulations later (Appendix 2 and 3), without going all the way to  the the bitter end.}

This integral equation can be solved iteratively, which we demonstrate in Section 7 below.

 When $\kappa=0$, the trading intensity is seen to be
\begin{equation}\label{E: AOB1}
\beta_t^{0} = \frac{\sigma_t^2\gamma_t(\beta)}{2\int_t^T\gamma_s(\beta)^2\beta_s ds},  \quad (\kappa = 0.)
\end{equation}
This can further be reduced to the following simple expression (see Aase et.al (2012a,b))
\begin{equation}\label{AOB}
\beta_t^{0} = \frac{\sigma_t^2\Big(\int_0^T\sigma_s^2ds\Big)^{\frac{1}{2}}}{\sigma_{\tilde{v}}\int_t^T\sigma_s^2ds},  \quad\text{when} \,\,\kappa = 0.
\end{equation}
When $\sigma_t = \sigma$ for all $t \in [0, T]$, where $\sigma$ is a positive constant, this  finally reduces to the Kyle (1985)-solution.

\section{The regulator's problem}

To limit the distortion of prices, a regulatory authority (the SEC) imposes an upper bound on price volatility. This is by and large the same as limiting the conditional expected degree of  price distortion (see Aase and Gjesdal (2018)).  In our model this limits the market maker's freedom to set prices. The market maker in our model is not really strategic, is risk neutral but exercises a certain degree of monopoly, as explained earlier. The regulator is introduced to mitigate this. 

As in the standard model, informed traders realize what the market maker is up to, and take his behavior into account when deciding their own trade. Noise traders just trade. In this situation the market maker can make unbounded profits taking advantage of noise traders, which would not make sense. To avoid this outcome the regulator is introduced.

It is well acknowledged that insider trading increases the volatility of an asset. Also price perturbations caused by the market maker's trading fees increase the volatility. This can be utilized by the regulator, who can suspend the stock from further trading based on  observing volatility over a certain acceptable, preset limit. A measure of volatility we consider as the basis for the regulator's ability to monitor the market.

The decision variable $\kappa$  of the market maker has so far "no dimension", meaning that it is  not an  observable quantity in the market. We therefore seek a connection between this variable and and an observable quantity. This is an important step in the analysis, because it allows us  to see if the market maker can really outperform  the  perfectly well informed insider in terms of profits at \emph{reasonable}  levels of trading fees, i.e., at levels where the regulator has not suspended the security.

From our expressions for the profit functions of the participants, we  notice that as $\kappa$ increases, the market maker's profit grows without limits, see e.g., Figure 2, and  eventually it will  dominate the profit function of the insider. The interesting question  is  then if this  takes place at an \emph{acceptable} level of price perturbations, set by the regulator.

With this in mind, we would like to develop a connection between the decision variable $\kappa$ and total volatility. Consider the quantity 
$\text{var}(p_t) = E(p_t- E(\tilde{v}))^2$. It is closely connected to the mean square deviation 
\[
S_t(\beta) = E(p_t - \tilde{v})^2 = \gamma_t(\beta) + k_t^2E(y_t^2).
\]
This latter quantity, or its square root, we assume can be observed by the regulator, who  will then compare this to the corresponding term $\gamma_t(\beta^0)$ based on no price distortions by the market maker. 

Recall the following definitions. $S_t(\beta) = E(\tilde{v} - p_t)^2$  and $\gamma_t(\beta^0) = E(\tilde{v} - m_t)^2$ where $m_t = E(\tilde{v}|\mathcal{F}_t^y)$.  The function $\gamma_t(\beta^0)$ corresponds to the expected square deviation between the true value of the asset and the fiducial price $m_t$, provided the trading intensity $\beta_t^0$ is used in the computation of the latter quantity. $S_t$ is the expected square deviation between the true value of the asset and the actual price that the market maker sets, in the  case where she trades  fees, as explained. Naturally, $S_t(\beta)$ is larger than $\gamma_t(\beta^0)$, and increasingly so as the market maker's decision variable $\kappa$  increases. 

This leads us to introduce  the following quantity in relative terms (rv = relative volatility)
\begin{equation}\label{E: d}
rv(t, \kappa) := \frac{\sqrt{S_t(\beta)}}{\sqrt{\gamma_t(\beta^0)}}, \quad t \in [0,T], \, \kappa \geq 0.
\end{equation}

Our assumption that $S_t : =S_t(\beta)$ is observable by the regulator also means that $rv(t, \kappa)$ is observable. 

From our previous results $S_t(\beta) = \gamma_t(\beta) + k_t^2V_t(k)$, where $V(t)$ depends on $k_t$ and is given by equation (\ref{E: eysquare}), which is
\begin{equation*}
V(t) = E(y_t^2) = e^{-2\int_0^tk_s\beta_sds}\int_0^t\sigma_s^2 e^{2\int_0^sk_r\beta_rdr}ds,
\end{equation*}
and  from (\ref{E: gamma}) we have that 
\begin{equation*}
\gamma_t(\beta^0) = \frac{\sigma_{\tilde{v}}^2}{1 + \sigma_{\tilde{v}}^2\int_0^t(\tilde{\beta}^0_s)^2ds},
\end{equation*}
where $\tilde{\beta}^0_t := \frac{\beta_t^0}{\sigma_t}$, and $\gamma_0 = E(\tilde{v} - E\tilde{v})^2 = \sigma_{\tilde{v}}^2$.
Using these relations, $rv(t, \kappa)$ can be written
\begin{equation}\label{E: d1}
rv(t, \kappa) =  \bigg(\frac{\gamma_t(\beta)}{\gamma_t(\beta^0)} + \frac{k_t^2}{\gamma_t(\beta^0)} e^{-2\int_0^tk_s\beta_sds}\int_0^t\sigma_s^2 e^{2\int_0^sk_r\beta_rdr}ds\bigg)^\frac{1}{2}.
\end{equation}
When $\kappa = 0$ we see from this expression that $rv(t, 0) \equiv 1$.

Here $rv$ will give information about the degree to which the market maker trades fees. When $\kappa = 0$, the function $rv(t, 0)$ is constant through time and identically equal to $1$ as noticed. As $\kappa$ increases from zero, $rv(t, \kappa)$ will rise above $1$,  and indicate the percent-wise increase from the situation with fiducial trade, at every $t \in [0,T]$.

For example, when $rv(t, \kappa) = 1.20$,  for some $t$ and $\kappa$, the actions of the market maker has increased the volatility of the asset by $20\%$ relative to the situation with fiducial price setting, where $\kappa = 0$. Thus the quantity $rv$ seems like a reasonable measure of  the degree of fee trading, in the hands of a regulator.

In this situation the map from $\kappa$ to $rv$, and in particular  its \emph{inverse} mapping, will  serve as a guide for acceptable values of $\kappa$, given a certain level of $rv$ set by the regulator.  This inverse mapping is illustrated in Section 7.5 below\footnote{In this section numerical illustrations can also be found, see in particular the two first rows in Table 2.}. The market maker can, on the other hand, use this mapping to exercise restraint in setting $\kappa$ in order to keep  markets open.

\section{A measure of price  informativeness}
We now derive a measure of the informativeness in prices, which is of particular interest when the prices are distorted.  Consider the quantity
\[
\iota(t, \kappa) := 1 - \frac{\text{var}(\tilde{v}|p_t)}{\text{var}(\tilde{v})}.
\]
When the price carries no private information about the true value of the asset at some time $t$ and some level of distortion $\kappa$, the conditional variance equals the unconditional variance, and consequently $\iota(t, \kappa) = 0$ at this point $(t, \kappa)$. When the price equals the value of the asset, the conditional variance equals $0$ and $\iota = 1$ at this point, in which case all the private information is reflected in the price. Consequently $0 \leq \iota(t, \kappa) \leq 1$ for all time  points $t \in [0,T]$ and for all $\kappa \geq 0$.

Because of the joint normal assumption, 
\[
\text{var}(\tilde{v}|p_t) = \text{var}(\tilde{v})(1 - \rho_{\tilde{v},p_t}^2)
\]
where $\rho_{\tilde{v},p_t}$ is the correlation coefficient between $\tilde{v}$ and $p_t$. Consequently $\iota(t, \kappa)  = \rho_{\tilde{v},p_t}^2$ for all $(t, \kappa)$.

In order to find this measure of price informativeness, we need to compute the  quantities cov$(\tilde{v}, p_t)$ and var$(p_t)$. To this end,  we first consider the covariance. Since $p_t = m_t + k_ty_t$,
\[
\text{cov}(\tilde{v}, p_t) = \text{cov}(\tilde{v}, m_t)  + k_t\text{cov}(\tilde{v}, y_t) .
\]
The first term on the right-hand side can be written 
\[
\text{cov}(\tilde{v}, m_t) = E(\tilde{v}m_t) - E(\tilde{v})E(m_t) = E(E(\tilde{v}m_t|\mathcal{F}_t^y)) - \mu_{\tilde{v}}^2 =
\]
\[
=E(m_tE(\tilde{v}|\mathcal{F}_t^y)) - \mu_{\tilde{v}}^2 =  E(m_t^2) - (E(m_t))^2 = \text{var}(m_t),
\]
since $m_t$ is $\mathcal{F}_t^y$-measurable, where $m_t = E(\tilde{v}|\mathcal{F}_t^y)$ and  $E(m_t) = \mu_{\tilde{v}}$. Furthermore, 
\[
\gamma_t(\beta) = E(\tilde{v} - m_t)^2 = E(\tilde{v}^2) - E(m_t^2) = \text{var}(\tilde{v}) - \text{var}(m_t),
\]
by a similar type of  conditioning as above. The last equality follows since $E(\tilde{v}) = E(m_t) = \mu_{\tilde{v}}$. Since we already have an expression for $\gamma_t = \gamma_t(\beta)$, see equation (\ref{E: gamma}), we now have an expression for $\text{var}(m_t)$, and hence  $\text{cov}(\tilde{v}, m_t) = \text{var}(m_t) = \sigma_{\tilde{v}}^2 - \gamma_t(\beta)$ by the above.

 The term $\text{cov}(\tilde{v}, y_t) $ is calculated as follows: First notice that by iterated expectations $\text{cov}(\tilde{v}, y_t)  = \text{cov}(m_t, y_t)$. From the Kalman filter equation we have
 \[
 dm_t = \frac{\beta_t\gamma_t(\beta}{\sigma_t^2}dy_t,
 \]
 see equation (3.9), and from the binormality between $m_t$ and $y_t$ and the corresponding projection theorem we obtain the following connection
 \[
 \rho_{m_t,y_t} = \frac{\beta_t\gamma_t(\beta)\sigma_{y_t}}{\sigma_t^2 \sigma_{m_t}},
 \]
 were  $ \rho_{m_t,y_t}$ is the correlation coefficient between $m_t$ and $y_t$, $\sigma_{y_t} := \sqrt{V(t)}$ and $\sigma_{m_t} := \sqrt{\text{var}(m_t)} = \sqrt{\sigma_{\tilde{v}}^2 - \gamma_t(\beta)}$.
 
 We have then shown that
 \[
 \text{cov}(\tilde{v}, p_t) = \sigma_{\tilde{v}}^2 - \gamma_t(\beta) + k_t\rho_{m_t, y_t} \sqrt{\sigma_{\tilde{v}}^2 - \gamma_t(\beta)}\sqrt{V(t)},
 \]
 where we have formulas for all the terms on the right-hand side of this equation.
 
 It remains to find the variance of $p_t$. Again, from $p_t = m_t + k_t y_t$ it follows that
 \begin{multline}\label{E: varp}
\text{ var}(p_t) = \text{var}(m_t) + k_t^2\text{var}(y_t) + 2k_t\text{cov}(m_t, y_t) =\\
\sigma_{\tilde{v}}^2 - \gamma_t(\beta) + k_t^2 V(t) + 2k_t\rho_{m_t,y_t} \sqrt{\sigma_{\tilde{v}}^2 - \gamma_t(\beta)}\sqrt{V(t)}.
\end{multline}
 Putting all this together, we have
 \begin{equation}\label{E: rho}
 \rho_{\tilde{v}, p_t} = \frac{\sigma_{\tilde{v}}^2 - \gamma_t(\beta) + k_t\rho_{m_t,y_t} \sqrt{\sigma_{\tilde{v}}^2 - \gamma_t(\beta)}\sqrt{V(t)}}{\sigma_{\tilde{v}}\sqrt{\text{var}(p_t)}}
\end{equation}
where var$(p_t)$ is given above in (\ref{E: varp}). From this the informativeness $\iota$ in prices is  given by
\begin{equation}\label{inform}
\iota(t, \kappa) = ( \rho_{\tilde{v}, p_t})^2,
\end{equation}
where $\rho_{\tilde{v}, p_t}$ is calculated using (\ref{E: rho}) and (\ref{E: varp}).

Table 1 illustrates the time development of the informativeness in the market. For a given value of the price distortion parameter $\kappa$ we notice that the informativeness increases with time, see the last row in Table 1.

 In the same table we have also computed some of the other key quantities that is used in the computation of the measure of informativeness $\iota(t, \kappa)$, such as the correlation coefficients $\rho_{m_t,y_t}(t, \kappa)$ and  $\rho_{\tilde{v}, p_t}(t, \kappa)$ and the variances of the price $p_t$ and  the 'fair' price $m_t$. While $\iota(t, \kappa) = \rho_{\tilde{v}, p_t}(t, \kappa)^2$ and thus $\rho_{\tilde{v}, p_t}(t, \kappa)$ represents an equivalent measure of information as $\iota(t, \kappa)$, the correlation coefficient $\rho_{m_t,y_t}(t, \kappa)$ is a measure connected to the 'fair' value $m_t$ instead of the market price $p_t$, but  computed with the value of $\beta_t$ where the insider has optimally adjusted to the actual distortion of the price. 
 
 This measure tells us how closely correlated the 'fair' price $m_t$ is to the order flow $y_t$. From the table we notice that  there can be a high correlation, as in our example, which throws some new light on the market maker's advantage   in observing the order flow.
 
 In the example this measure decreases with time up to a certain value $t^*$, then increases after that, so that here the advantage is highest in the beginning and towards the end of the trading interval.
 
  Since the correlation cov$(\tilde{v}, m_t) =$\,var$(m_t)$, this covariance also  increases  slowly with time in our example, which is reasonable since  more information about the true value of the asset becomes available with time, hence the increase in the corresponding correlation coefficient $\rho_{\tilde{v},m_t}= \sqrt{\text{cov}(\tilde{v}, m_t)}/\sigma_{\tilde{v}}$. At the same time this shows that $\gamma_t(\beta)$ decreases slowly with time, again for the same reason: One knows more about the true value as time increases, and the present  measure of uncertainty  (the mean square error) then naturally decreases.

Finally, we illustrate the time development of the  variance of the price $p_t$ and of the 'fair' price $m_t$. The latter one has  just been explained. For the former two different effects are in force: One is  that $V(t)$ increases with time (see Section 7.3 below) and what has just been shown, that cov$(\tilde{v}, m_t)$ also increases with time. The other is that $k_t$ decreases with time,  see  the expression (\ref{E: varp}) for the var($p_t$). In our example the first effect weakly dominates.

 \begin{table}[h]
\centering
\begin{tabular}{l|c|c|c|c|c|c|c|c|c|c}
\hline
\hline
Cont.  model:		 \\
\hline
$t$      		                          &     1     &        2   &    3     &   4      &   5   &            6    &         7   &              8         &   9          &   10\\ 
\hline
\hline
$\rho_{m_t,y_t}(t, \kappa)$    &   .96      &    .92   &   .87      &   .84       &   .81          &  .79   &          .78            &   .79   &  .83      &     1.00  \\
\hline   
$\rho_{\tilde{v}, p_t}(t, \kappa)$ &   $.31$  &    .44  &   .53	     & .61    &  .67        &    .73 	  &	 .79       &   .84	      & .91    &       .99  \\
\hline   
$\text{var}(p_t)$             &.02          & 	  .04    &    .05	     & .06    &  .07      &  .07  &	.08      & .08	 & .08     &       .09   \\
\hline
$\text{var}(m_t)$             &.009          & 	  .02    &    .03	    & .03     &  .04      &  .05  &	.06      & .06 	 & .07     &       .09   \\
\hline
$\iota(t, \kappa)$              &  .10             &  .19  &     .28        &   .37        &  .45       &   .54          &     .62        & .71       & .83         & 1.00 \\    
\hline
\hline
\end{tabular}
\caption{The quantity $\iota(t,\kappa)$ as a function of time  ($\kappa = 0.035$).}
\label{tbl: two profits}
\end{table}

The shape of  $\iota(t, \kappa)$ as a function of  the price distortion parameter $\kappa$  for a given value of time $t$ is illustrated in the next section, see  the last row of Table 2 below. Naturally we expect that  $\iota(t, \kappa)$ decreases with $\kappa$ for any given value of $t$.

\section{Illustrations of the theoretical results}
\subsection{General}
Based on the integral equation (\ref{E: opt beta}) for  trading intensity of the insider, we now  present a few illustrations. 

First, we  indicate how to deal with this integral equation. This equation can be  transformed to a differential-integral equation if one so pleases, but we choose to work with the version we have, where we use an iterative procedure.

As a first step we suggest to  use a  trial solution, $\beta_t^0$ say,  on  the right-hand side of this equation, and then find  the first approximation, $\beta_t^1$,  given by the left hand side as a function of this initial solution. A reasonable candidate for the trial solution is of course the solution when $\kappa = 0$, which we have in closed form (see (\ref{AOB})). Next one  continues this procedure, where $\beta_1(t)$ becomes the new trial solution in the next step, and so on until convergence.  This, of course, requires  the right numerical tools.

\subsection{The trading intensity of the insider}

In Figure 3 below we illustrate the time development of  the trading intensities of the insider obtained this way, where the upper curve is when $\kappa=0$ and the lower curve is for $\kappa = 0.045$. The other base case parameters are the same as in Figure 1. We choose $T = 10$ here as well\footnote{Here and in most of the numerical computations we do not go beyond two rounds in the iterations indicated above.}.\\

\vbox{\centerline{\includegraphics[width=45mm]{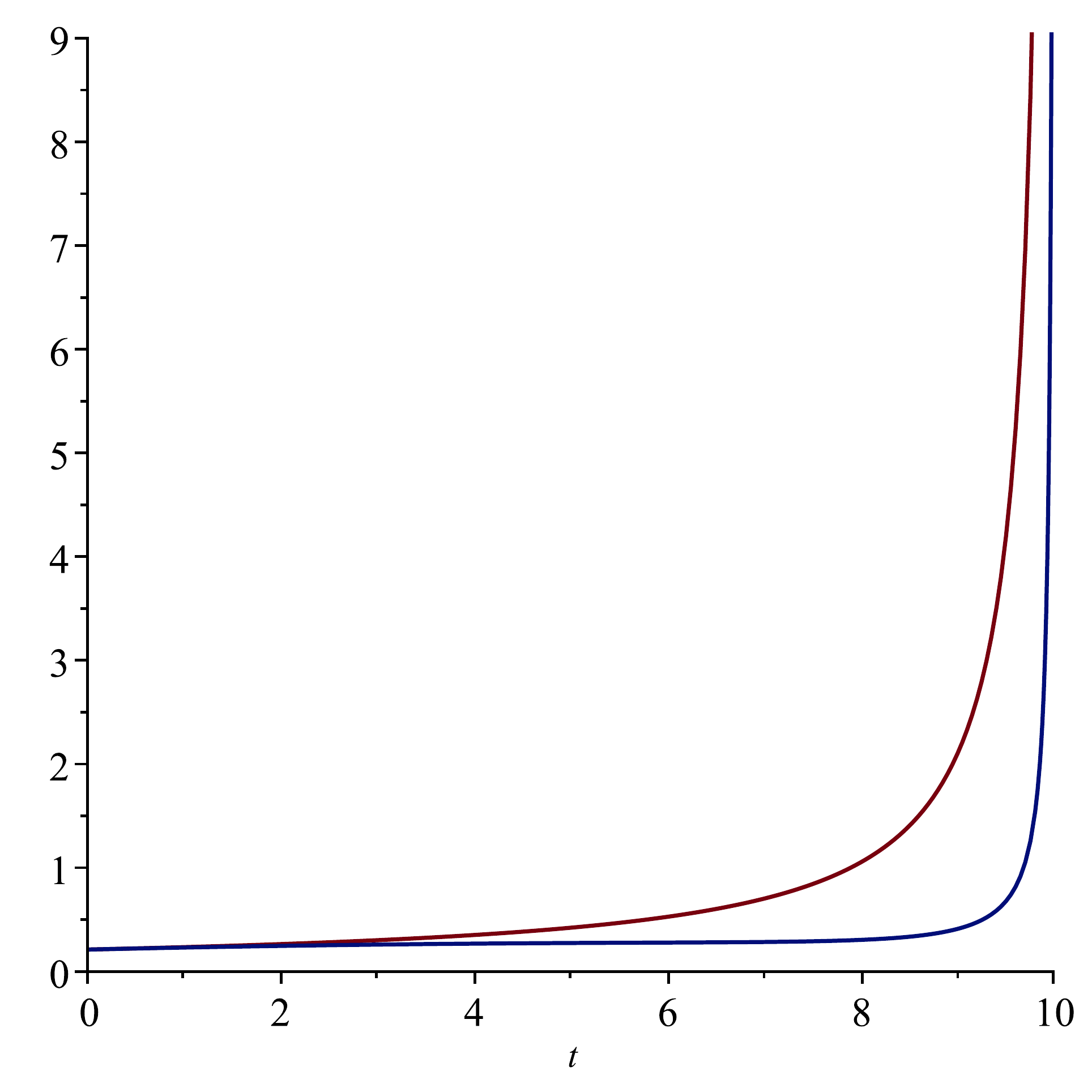}}
\centerline{Fig. 3: The insider's trading intensities  as functions of $t$.}}

Note that in both cases the insider intensifies her trade towards the horizon. Also,  it is reasonable that the lower graph corresponds to the insider's trading intensity when the market maker perturbs the price. The insider, knowing what the market maker is up to, now trades more softly. However, towards the end she picks up trading again, since then the market maker trades fees  to a  less and less extent as $t$ approaches $T$.

This type of analysis represents an interesting extension of the analysis in Aase and Gjesdal (2018), who  treat the one period model. In that model the intensity is graphically represented as a decreasing, convex function of the decision variable  $\kappa$ (but there is no time development). \\

\vbox{\centerline{\includegraphics[width=45mm]{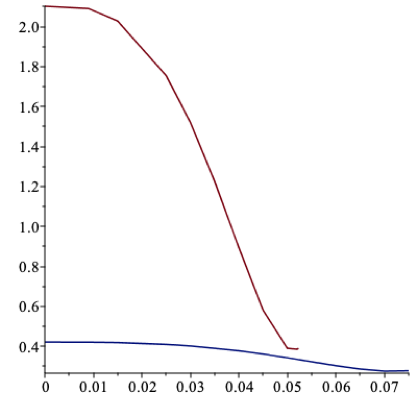}}
\centerline{Fig. 4: The insider's trading intensities as functions of $\kappa$.}}
 
In Figure 4 we present two graphs of $\beta(t, \kappa)$ in the present continuous model as a function of $\kappa$ for  given $t$, for two different points in time: $t =  5\, \,\text{and}\, \, t = 9$, where the upper curve is for $t = 9$.  These are also seen to be decreasing.  Unlike for the one-period model, these graphs are concave for small values of $\kappa$ and then becomes convex as $\kappa$ increases. That these graphs are decreasing in $\kappa$ is in accordance with the results from the  one period model: The insider trades more softly the more the market maker perturbs the price.

\subsection{The variance function of the order flow y}

We have already indicated a graph,  Figure 1, illustrating the covariance function of the order flow. Here we study the time development of the variance function $V(t)$, as this quantity enters many of the key expressions in this theory. The base case parameters are the same as before.\\

\vbox{\centerline{\includegraphics[width=45mm]{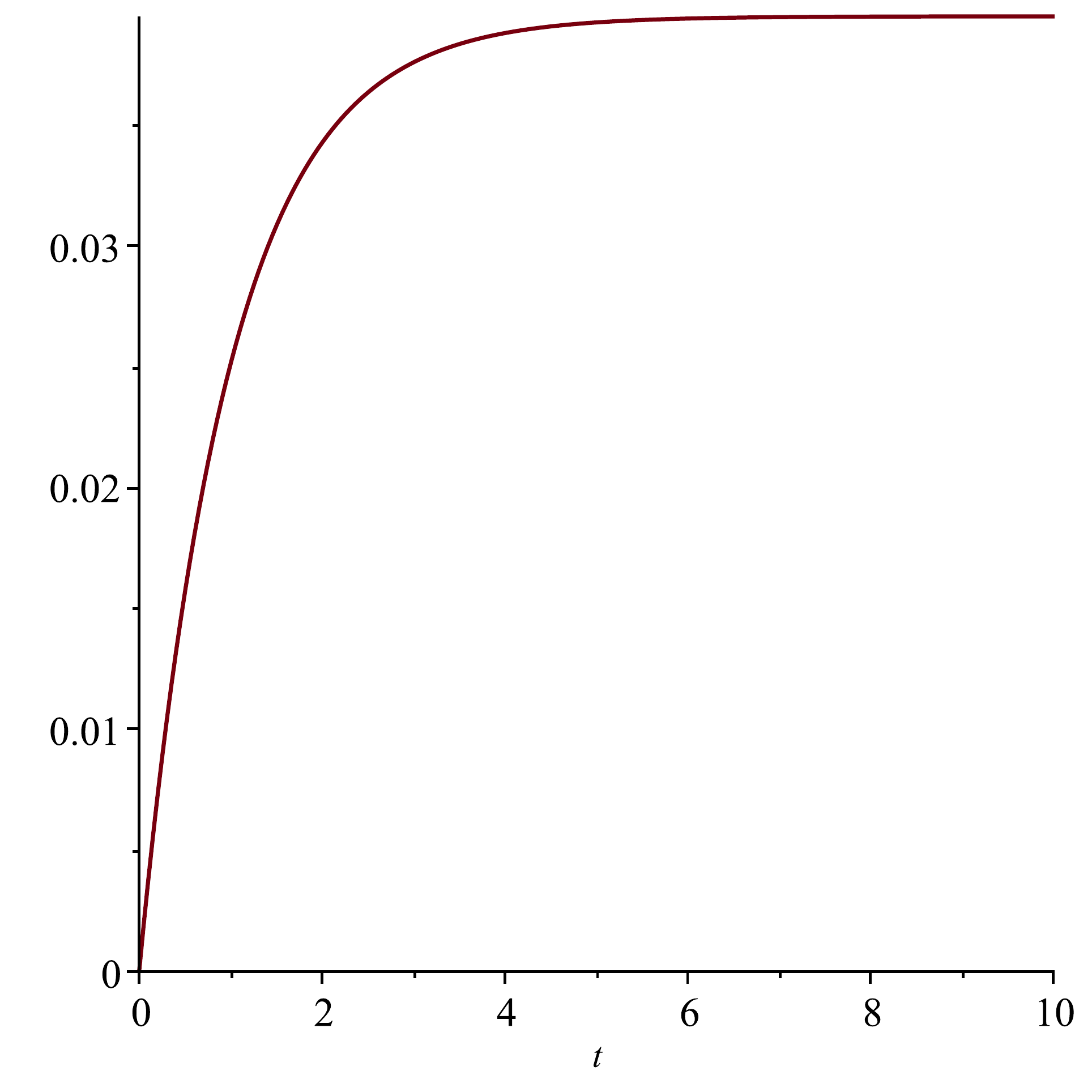}}
\centerline{Fig. 5: The variance function $V(t, \kappa)$ as a function of  $t$. }}

The variance of the total order flow $y_t$ is seen to increase sharply in the beginning, and then flattens out  at around $t = 3$ when $\kappa = 0.24$.  It is well known from  empirical studies that the  volatility of the price increases as more relevant information  enters the market, since this causes trade to increase. A reasonable model of insider trading should reflect this, and for our model this is  illustrated in Figure 5.

However, note that these variances decrease with $\kappa$ for any given point $t$ in time. This is illustrated in Figure 6 for $t = 1, 5, 9$ as $\kappa$ run from $0$ to $0.5$. The more the market maker trades fees, the less the insider trades and the lower the variance of the order flow $y_t$.\\

\vbox{\centerline{\includegraphics[width=45mm]{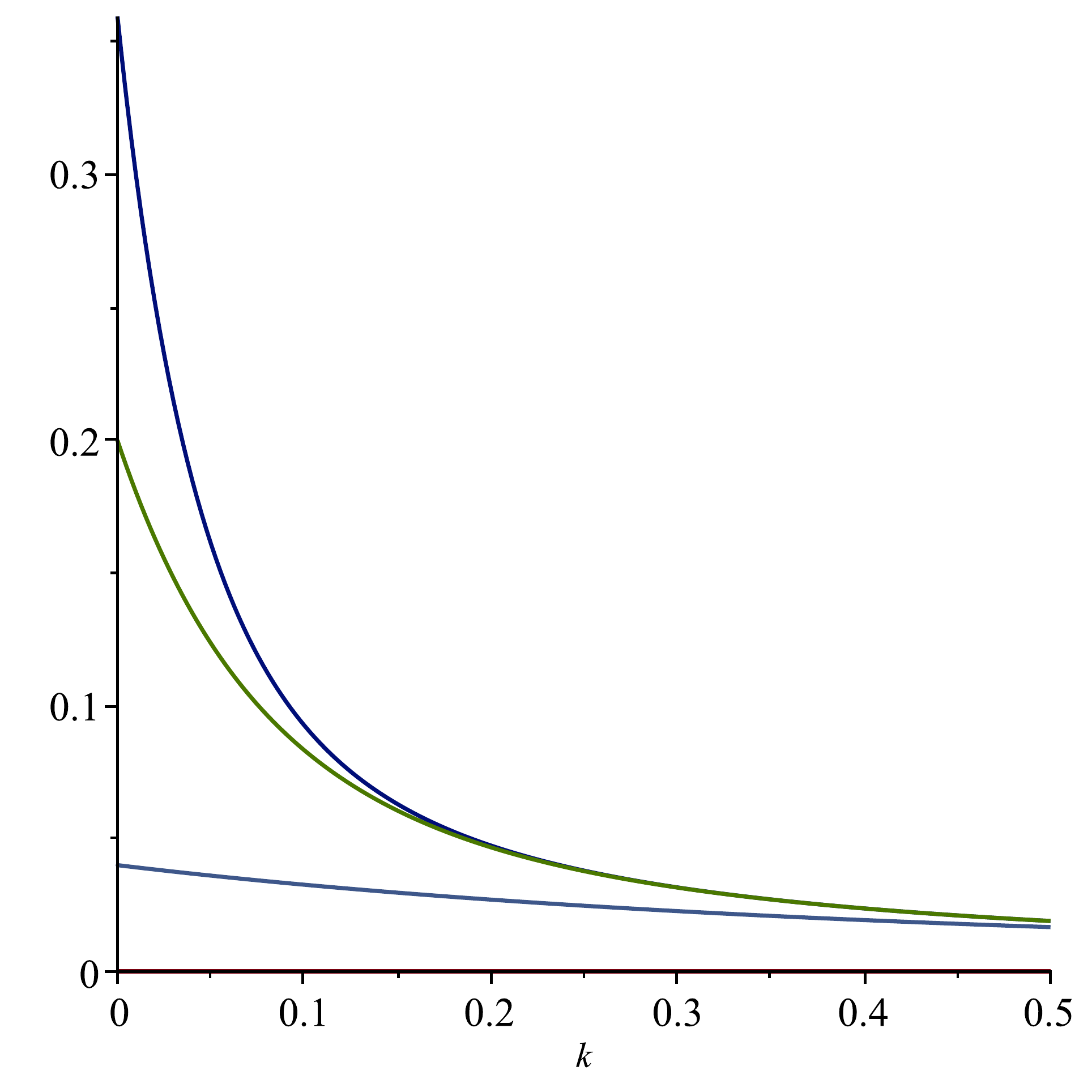}}
\centerline{Fig. 6: The variance function $V(t, \kappa)$ as a function of  $\kappa$.}}

The upper curve  in Figure 6 corresponds to $t = 9$, then $t = 5$ and the lowest curve is for $t =1$.

\subsection{The  two parties' net profits}

Now we come to the important part, namely the net profit functions of the two key participants as functions of time.  These are illustrated in figures 7 and 8. 

In Figure 7 we consider the dynamics of the two net profits  as functions of  of time, for a given value of the parameter $\kappa$. We notice that both profit functions  naturally start out low and then increase with time. When $\kappa = 0.045$ the insider's profit  function  cross the market maker's profit from below at around $t = 7$, and then ends up as the highest of the two at the final time  $T = 10$\footnote{The profit function of the insider is  computed at discrete times only, due to the large number of computations required for a continuous plot.} . \\

\vbox{\centerline{\includegraphics[width=45mm]{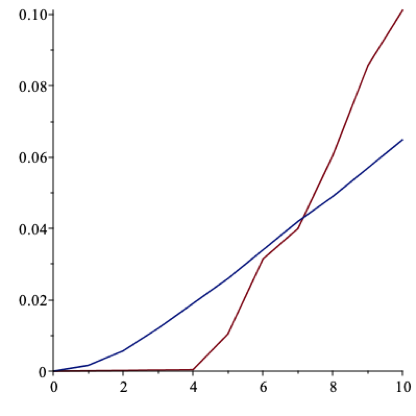}}
\centerline{Fig. 7: The two net profits as functions of time. $\kappa = 0.045$.}}

In Figure 8 however,   where $\kappa = 0.07$,  the market maker's profit  is seen to dominate from the start, and also ends up highest at  the final time $T = 10$. When the market maker further increases $\kappa$, her profit function will  increase for each value of $t$, while the profit of the insider will decrease for each $t$ compared to the levels in Figure 8, (but will still be an increasing function of $t$ for each given value of $\kappa$, as in the figures 7 and 8). As a consequence, the market maker outperforms the perfectly well informed insider from about   $\kappa = 0.06$ onwards.\\

\vbox{\centerline{\includegraphics[width=45mm]{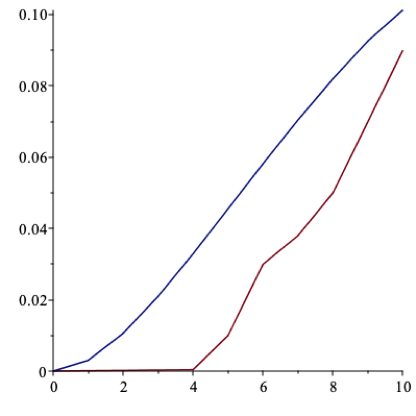}}
\centerline{Fig. 8: The two net profits as functions of time. $\kappa = 0.07$.}}

\subsection{When does the market maker's profit dominate?}
We now illustrate the regulator's problem through some graphs of the market observable quantity $rv$.  In the first figure we plot $rv(t, \kappa)$ as functions of $\kappa$ for some given values of time $t$.\\

\vbox{\centerline{\includegraphics[width=45mm]{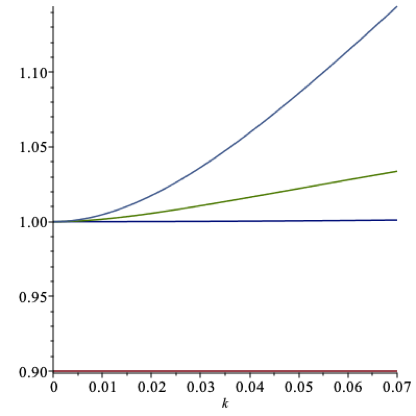}}
\centerline{Fig. 9: $rv(t,\kappa)$ as a function of  $\kappa$ when $t = 0.01, 3.6$ and $9.0.$}}

The lowest curve in Figure 9 corresponds to $t = 0.01$, the next lowest to $t = 9.0$ and the highest curve to $t = 3.6$. To interpret  this figure, imagine that the regulator uses the rule that $rv$  should correspond to  less than  $15\%$ increase from the fiduciary, ideal situation. This means that the  market maker should not increase $\kappa$  beyond $0.07$, at least based on the  the three time  points in this illustration. 

Since this line of reasoning is valid only for some particular values of $t$, in the next figure we study $rv$'s time development:\\

\vbox{\centerline{\includegraphics[width=45mm]{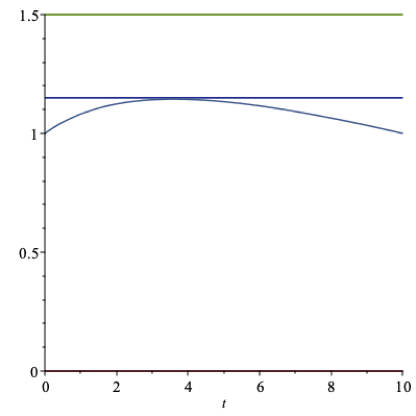}}
\centerline{Fig. 10: $rv(t,\kappa)$ as a function of  $t$ when $\kappa= 0.07.$}}

Figure 10  illustrates the  time picture of $rv(t,\kappa)$ for $\kappa = 0.07$, for $t \in [0,T]$ (the curved graph). The horizontal line tangent to the curve $rv(t, 0.07)$ is at level $1.15$, corresponding to $15\%$ maximal deviation from fiduciary trade. From the figure it follows that the regulator will keep the market open  all the  time.\\

\vbox{\centerline{\includegraphics[width=45mm]{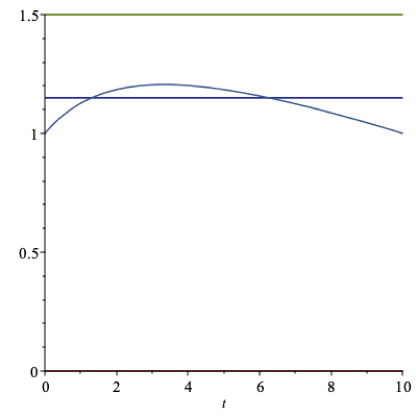}}
\centerline{Fig. 11: $rv(t,\kappa)$ as a function of  $t$ when $\kappa= 0.09.$}}

Figure 11  illustrates the same  time picture of $rv(t,\kappa)$  as in the previous figure, but now for  $\kappa = 0.09$. 
With a  $15\%$-regulation rule still in charge,  the market would then be shut down and the stock suspended  at around $t = 1.2$, but allowed to reopen at  around $t = 6.5$, and then kept open for the rest of the period.  Accordingly, the market maker would be wise to consider a lower value of $\kappa$ to keep her reputation as a decent professional.
 
On the other hand, if the regulator used a $21\%$-rule, the market would stay open all the time for  $\kappa = 0.09$.

The manner in which we find the \emph{inverse} map from $rv$ to $\kappa$ is seen to proceed as follows: For a given value of $rv^*$, say  $rv^* = 1.15$, we solve in $t$ and $\kappa$ the following inequality: $\text{sup}_{\kappa}(\text{sup}_t rv(t, \kappa))\leq d^*$. Then the 'optimal' value of $\kappa$, call it $\kappa^*$, is the one which satisfies this inequality with equality sign:  $\text{sup}_t rv(t, \kappa^*) = d^*$.  In Table 2 we illustrate this connection for some values of $rv^*$.

  \begin{table}[h]
\centering
\begin{tabular}{l|c|c|c|c|c}
\hline
\hline
Cont.  model:	$t=9$	  \\
\hline
$rv^*$      		                          &     1.03      &         1.05    &      1.08     &   1.15       &   1.21\\    
\hline
\hline
$\kappa^*$                                 &      .025        &        .035      &   .045     	&   .070  	&     .090   \\ 
\hline   
$p_M(9, \kappa^*)$                      &    .037 	&	.049     	& .060 	& .087     &       .100     \\
\hline   
$p_I(9, \kappa^*)$             	     &.126 		&	.118		& .100 	& .055   &       .045     \\
\hline
$\iota(9, \kappa^*)$                        &   .88      	&       .83       &   .64    	&   .45       &      .30     \\    
\hline
\hline
\end{tabular}
\caption{The connection between $\kappa^*$and  $rv^*$ with associated net profits.}
\label{tbl: two profits}
\end{table}
In  Table 2 the insider has the highest net profits, denoted  $p_I(t, \kappa)$,  for lower values of $rv^*$, and this profit is decreasing with $k^*$. For larger values of $rv^*$, the market maker's profit is the largest of the two. The last row in Table 2 illustrates the informativeness  $\iota(t, \kappa)$ in the market as a function of $\kappa$ when $t = 9$. It decreases as $\kappa$ increases. Distorting prices is not informative to the other market participants.

In the above illustrations  in the last two figures (figures 10 and 11) it is  the market maker  who has the highest profits of the two parties. This highlights one of the the main ideas in this paper: In real life we know that market makers actually do not set prices in an entirely fiduciary manner, but rather determine prices in such  a way that they make money. In the introduction we explained why this behavior is  possible  and  likely to take place. This is in line with observed behavior in several financial markets, in particular those  of the over-the-counter type that we have in mind.  

For a modest fee conditional on the order flow, the market maker is able to obtain a profit of the order of the magnitude, or even better than, a perfectly informed insider, showing, among other things, the advantage of observing the order flow. This,  we conjecture, may be one explanation why so much money tends to end up in the financial sector of the economy.

\section{Suggestions for further research}
In the discrete time paper of Aase and Gjesdal (2017) a situation  is analyzed where the market maker has private information as well.  This could also be of interest to analyze in the present setting. Thee is supposed to be no information  flow between the market making  department and the investment department of large financial institutions. But these 'Chinese Walls' - as they are known as - may not be entirely 'watertight'.  Compared to the model where the market maker has no privileged information,  several new insights were obtained in the above mentioned paper. As expected the well informed insider's profit diminishes since his informational advantage has shrunk.  This is because the market maker acts non-strategically with respect to his private information, and allows this  information to be reflected into the price. Interestingly the informed market maker may then increase his fees without being detected by the regulator, which in turn increases his profits.

Finally, we could analyze the situation where the market maker's information is public, which can be used to determine the effects of information asymmetry. One major difference,  discovered in the above mentioned paper, is that the insider then trades harder when information is public, increasing his own profit and that of the market maker. As another consequence, the market is more efficient. Price distortions are now smaller as a function of the price perturbation parameter allowing the market maker to make higher profits. Thus it pays for him to share his private information with the insider, but the noise traders are then even worse off.  This analysis also throws some new light on one 'positive' side of insider trading (aside from the obvious negative aspects which we do not dwell on here); all information arriving to the market generally has a positive effect on efficiency.

In particular, price volatility is shown to increase with informed market maker in the one period model. This is an important aspect of  the effect of  privileged information on security prices, which may explain the price/dividend puzzle, a feature that should be extend to the multiperiod model (the world is, after all, time continuous).

\section{Conclusions}
The dynamic auction model of Kyle (1985) is studied, allowing market makers to maximize profit within regulatory limits by charging time varying, stochastic fees. This has several implications for the equilibrium, the most important  being that  by perturbing the price by a relatively modest amount, the market maker is able to obtain a profit of the order of magnitude, and even higher than,  a perfectly informed insider. 

 The dynamic aspects of this are analyzed in the paper, and illustrated numerically by examples. The analytical  challenge turned out to be the determination of the optimal trading intensity of the insider, when the market maker perturbs the price. The solution was presented in Theorem 1 in Section 4. Based on this result, we were able to discuss a wide variety of problems, like finding the profits of the two parties, the insider and the market maker, the stochastic properties of the order flow, and the informativeness in the market, all quantities as a functions of time and price perturbations. We also indicated how a regulator can monitor the market by observing a  measure of  relative price volatility, which compares to the corresponding measure with fiducial price setting. This gives a convenient connection between price volatility and price perturbation.
   
 Our analysis  indicates why speculative prices are more volatile than predicted by fundamentals.

\section{Appendix 1.\\
 Proof of Theorem 1 by Variational Calculus.} 
 
We now want to  solve problem (\ref{P:1}) by use of  directional derivatives, or calculus of variations. Towards this end, let $\mathcal{A}$ be the family of all continuously differentiable functions $\beta : [0, T] \to R$ such that
\[
\int_0^t\frac{\beta_s^2}{\sigma_s^2} ds < \infty \quad \text{for all} \quad t < T.
\]

By this method we choose  an arbitrary function $\xi_t \in \mathcal{A}$, a sufficiently rich set in this regard, and define the real function $g$ by
\[
g(x) = J^I(k, \beta + x\xi);\quad x\in R.
\]
Then, assuming $\beta$ is optimal, $g$ is maximal at $x = 0$ and hence the first order condition of maximality at $\beta$ is
\[
0 = g'(0) = \frac{d}{dx}J^I(k, \beta + x\xi)|_{x=0} =
\]
\[
\frac{d}{dx}\bigg(\sigma_{\tilde{v}}^2\int_0^T \Big(1 + \sigma_{\tilde{v}}^2\int_0^t\frac{(\beta_s + x\xi_s)^2}{\sigma_s^2}ds\Big)^{-1}(\beta_t + x\xi_t)dt\bigg)|_{x = 0} \,+
\]
\[
\frac{d}{dx}\bigg(\int_0^T \Big((\beta_t  + x\xi_t) k_t^2 e^{-2\int_0^tk_s(\beta_s + x \xi_s)ds}\int_0^t\sigma_s^2 e^{2\int_0^sk_r(\beta_r + x \xi_r)dr}ds\Big)dt\bigg)|_{x = 0} \, =
\]
\begin{multline}\label{E: foc}
\int_0^T\big(\gamma_t(\beta) - 2\int_t^T\gamma_s^2(\beta) \beta_sds\big)\frac{\beta_t}{\sigma_t^2}\, \xi_t \, dt +\\ \int_0^T\big(k_t^2e^{-2\int_0^tk_s\beta_sds}\int_0^t\sigma_s^2e^{2\int_0^s\,k_r\beta_rdr}ds\big)\xi_t\, dt + \\
\int_0^T\big(\beta_tk_t^2e^{-2\int_0^tk_r\beta_rdr}(-2\int_0^t\, k_s\xi_sds)\int_0^t\sigma_r^2e^{2\int_0^rk_u\beta_udu}dr)dt + \\
\int_0^T\big(\beta_tk_t^2e^{-2\int_0^tk_r\beta_rdr}\int_0^t\sigma_s^2e^{2\int_0^sk_r\beta_rdr}(2\int_0^s\, k_r\xi_rdr)ds)dt = 0,\quad \forall \xi \in \mathcal{A}.
\end{multline}
The  second line on the left-hand side of (\ref{E: foc}) follows just as in Aase et. al (2012c), which presents a simple proof of the case $k=0$.

Consider the last two lines, and start with the third integral  in (\ref{E: foc}). By changing the order of integration between s and t, we obtain 
\[
\int_0^T\big(\beta_tk_t^2e^{-2\int_0^tk_r\beta_rdr}(-2\int_0^t\, k_s\xi_sds)\int_0^t\sigma_r^2e^{2\int_0^rk_u\beta_udu}dr)dt =
\]
\[
-2\int_0^T\int_s^T\big(\beta_tk_t^2e^{-2\int_0^tk_r\beta_rdr}(\int_0^t\sigma_r^2e^{2\int_0^rk_u\beta_udu}dr)dt\big)k_s\xi_sds = 
\]
\[
-2\int_0^T\int_t^T\big(\beta_sk_s^2e^{-2\int_0^sk_r\beta_rdr}(\int_0^s\sigma_r^2e^{2\int_0^rk_u\beta_udu}dr)ds\big)k_t\xi_tdt. 
\]
Next consider the fourth and last integral in  (\ref{E: foc}). The inner integral given by
\[
\int_0^t\sigma_s^2e^{2\int_0^sk_r\beta_rdr}(2\int_0^s\, k_r\xi_rdr)ds)dt,
\]
can be rewritten as
\[
2\int_0^t(\int_s^t\sigma_r^2e^{2\int_0^rk_u\beta_udu}dr) k_s\xi_sds.
\]
This we now insert in the fourth term in (\ref{E: foc}), which gives
\[
2\int_0^T\int_t^T\big(\beta_sk_s^2e^{-\int_0^tk_u\beta_udu}(\int_t^s\sigma_r^2e^{2\int_0^rk_u\beta_udu}dr)ds)k_t\xi_tdt.
\]
Putting all this together, the first order condition now takes the form
\begin{multline}
0 = \frac{d}{dx}J^I(k, \beta + x\xi)|_{x=0} = \\ \int_0^T\Big(\gamma_t(\beta) - 2(\int_t^T\gamma_s(\beta)^2\beta_sds)\frac{\beta_t}{\sigma_t} + 
k_t^2e^{-2\int_0^tk_s\beta_sds}\int_0^t\sigma_s^2e^{-2\int_0^tk_r\beta_rdr}ds \\ - 2k_t\int_t^T\beta_sk_s^2e^{-2\int_0^sk_r\beta_rdr}(\int_0^s\sigma_r^2e^{2\int_0^rk_u\beta_udu}dr)ds\\ + 2k_t\int_t^T\beta_sk_s^2e^{-2\int_0^tk_u\beta_udu}(\int_t^s\sigma_r^2e^{2\int_0^rk_u\beta_udu}dr)ds\Big)\xi_tdt,\quad \forall \xi \in \mathcal{A}.
\end{multline}
Thus we conclude that
\begin{multline}
\gamma_t(\beta) = 2\frac{\beta_t}{\sigma_t}\int_t^T\gamma_s(\beta)^2\beta_sds + 
k_t^2e^{-2\int_0^tk_s\beta_sds}\int_0^t\sigma_s^2e^{-2\int_0^tk_r\beta_rdr}ds \\ - 2k_t\int_t^T\beta_sk_s^2e^{-2\int_0^sk_r\beta_rdr}(\int_0^s\sigma_r^2e^{2\int_0^rk_u\beta_udu}dr)ds\\ + 2k_t\int_t^T\beta_sk_s^2e^{-2\int_0^tk_u\beta_udu}(\int_t^s\sigma_r^2e^{2\int_0^rk_u\beta_udu}dr)ds.
\end{multline}
Accordingly
\begin{multline}\label{E: opt beta 1}
\beta_t = \frac{\sigma_t^2}{2\int_t^T\gamma_s(\beta)^2\beta_s ds}\Big(\gamma_t(\beta) - k_t^2e^{-2\int_0^tk_s\beta_sds}\int_0^t\sigma_s^2e^{-2\int_0^tk_r\beta_rdr}ds \\ 
- 2k_t\int_t^T\beta_sk_s^2e^{-2\int_0^sk_r\beta_rdr}(\int_0^t\sigma_r^2e^{2\int_0^rk_u\beta_udu}dr)ds\Big).
\end{multline}
Hence, the optimal trading intensity of the insider, $\beta_t, t \in [0,T]$, is given by the integral equation  (\ref{E: opt beta 1}).

This may be simplified by using the expression for the variance process $V(t)$ in (\ref{E: eysquare}). The result is

\begin{equation}\label{E: opt beta0}
\beta_t = \frac{\sigma_t^2}{2\int_t^T\gamma_s(\beta)^2\beta_s ds}\Big(\gamma_t(\beta) - V(t)\big(k_t^2+
2k_t\int_t^T\beta_sk_s^2e^{-2\int_t^sk_r\beta_rdr}ds\big)\Big).
\end{equation}
which is equation(\ref{E: opt beta}) in Theorem 1. $\qquad \square$

\section{Appendix 2.\\
Optimization via Pontryagin and Nash.}

We start from  the system expressed in terms of $(y_t,m_t,\gamma_t)$ in (3.8)-(3.10), which are
\begin{align}
&dy_t=(\tilde{v}-m_t-k_ty_t)\beta_t dt +\sigma_t dB_t;\quad y_0=0 \\
&dm_t=\frac{\gamma_t\beta_t}{\sigma_t^2}[(\tilde{v}-m_t)\beta_t dt + \sigma_t dB_t];\quad m_0=E[v]\\
&d\gamma_t=- \frac{\beta_t^2 \gamma_t^2}{\sigma_t^2}; \quad \gamma_0=E[(\tilde{v}-E[\tilde{v}])^2].
\end{align}

The performance functionals are
\begin{align}
&J^{M}(k,\beta):=w_0^M + E(\int_0^T\, k_t y_t\,( k_ty_t  + m_t - \tilde{v})\beta_t dt - \int_0^Ty_t^2\, dk_t)\\
&J^{I}(k,\beta):= w_0^{I}+ \int_0^T E[(\tilde v -m_s - k_sy_s)^2]\beta_s ds.
\end{align}

\begin{problem} We want to find a Nash equilibrium $(k_t^*,\beta_t^*)$ for the the two performance functionals $J^{M}, J^{I}$.
In other words, we want to find (deterministic) control processes $k_t^*, \beta_t^*$ such that
\begin{equation}
\sup_{k_t}J^{M}(k_t,\beta_t^*) = J^{M}(k_t^*,\beta_t^*)
\end{equation}
and
\begin{equation}
\sup_{\beta_t} J^{I}(k_t^*,\beta_t) = J^{I}(k_t^*,\beta_t^*).
\end{equation}
\end{problem}
This is a \emph{stochastic differential game}. 


Recall our assumption
\begin{equation}
k_t=\kappa (T-t);\quad 0 \leq t \leq T,
\end{equation}
for some constant $\kappa \in [0,K],$ 
where $K$ is a fixed constant (in principal set by the regulator).

Then the performance functionals take the following forms
\begin{align}
&J^{M}(\kappa,\beta):=w_0^M + E[\int_0^T\{ \kappa(T-t) y_t\,( \kappa(T-t)y_t  + m_t - \tilde{v})\beta_t  + \kappa y_t^2\}dt]\\
&J^{I}(\kappa,\beta):= w_0 + \int_0^T E[(\tilde v -m_t - \kappa (T-t)y_t)^2]\beta_t dt.
\end{align}
To study Problem 3.1 we use the stochastic maximum principle. Thus we define two \emph{Hamiltonians} $H^M$ and $H^{I}$ by
\begin{align}
H^M(t,y,m,\gamma,\kappa,\beta,p,q)&=-\kappa(T-t)y(v-m-\kappa(T-t)y)\beta+\kappa y^2\nonumber\\
&+(v-m-\kappa(T-t)y)\beta p_1+\sigma_t q_1 + \frac{\gamma\beta}{\sigma_t^2}(v-m)\beta p_2\nonumber\\
&+\frac{\gamma\beta}{\sigma_t}q_2-\frac{\beta^2\gamma^2}{\sigma_t^2}p_3
\end{align}
and
\begin{align}
H^{I}(t,y,m,\gamma,\kappa,\beta,p,q)&=(v-m-\kappa(T-t)y)^2 \beta + (v-m-\kappa(T-t)y)\beta p_1  \nonumber\\
&+\sigma_t q_1+\frac{\gamma\beta}{\sigma_t^2} (v-m)\beta p_2 + \frac{\gamma \beta}{\sigma_t} q_2 -\frac{\beta^2 \gamma^2}{\sigma_t ^2} p_3.
\end{align}
The BSDE's for the adjoint processes $(p_i^M,q_i^M); i=1,2,3,$ associated to $H^M$ are
\begin{align}
\begin{cases}
dp_1^{M}(t)&=[\kappa(T-t)\beta_t(v-m_t-2\kappa(T-t)y_t) - \kappa^2(T-t)^2y_t\beta_t +\nonumber\\
& 2 \kappa y_t +\kappa(T-t)\beta_t p_1^{M}(t)]dt + q_1^{M}(t)dB_t;\quad 0\leq t\leq T\\
p_1^{M}(T)&=0
\end{cases}
\end{align}

\begin{align}
\begin{cases}
dp_2^M(t)&=-[-\kappa(T-t)y_t\beta_t-\beta_t p_1^M(t)-\frac{\gamma_t\beta_t^2}{\sigma_t^2}p_2^M(t)]dt+q_2^M(t)dB_t; 0 \leq t \leq T\nonumber\\
p_2^M(T)&=0
\end{cases}
\end{align}

\begin{align}
\begin{cases}
dp_3^M(t)&=-[\frac{\beta_t^2(v-m_t)}{\sigma_t^2} p_2^M(t) +\frac{\beta_t}{\sigma_t} q_2^M(t) - \frac{2\beta_t^2 \gamma_t}{\sigma_t^2} p_3^M(t)]dt+q_3^M(t)dB_t; 0 \leq t \leq T\nonumber\\
p_3^M(T)&=0
\end{cases}
\end{align}

The BSDE's for the adjoint processes $(p_i^{I},q_i^{I}); i=1,2,3,$ associated to $H^{I}$ are
\begin{align}
\begin{cases}
dp_1^{I}(t)&=-[2(v-m_t-\kappa(T-t)y_t)(-\kappa(T-t))\beta_t -\kappa(T-t)\beta_t p_1^{I}(t)]dt\nonumber\\
& + q_1^{I}(t)dB_t;\quad 0\leq t\leq T\\
p_1^{I}(T)&=0
\end{cases}
\end{align}

\begin{align}
\begin{cases}
dp_2^{I}(t)&=-[2(v-m_t-\kappa(T-t)y_t)(-\beta_t) - \beta_t p_1^{I}(t)-\frac{\gamma_t\beta_t^2}{\sigma_t^2} p_2^{I}(t)]dt\nonumber\\
&+q_2^{I}(t)dB_t; 0 \leq t \leq T\\
p_2^{I}(T)&=0
\end{cases}
\end{align}

\begin{align}
\begin{cases}
dp_3^{I}(t)&=-[\frac{v-m_t}{\sigma_t^2} \beta_t^2 p_2^{I}(t) +\frac{\beta_t}{\sigma_t} q_2^{I}(t) - \frac{2\beta_t^2 \gamma_t}{\sigma_t^2} p_3^{I}(t)]dt+q_3^{I}(t)dB_t; 0 \leq t \leq T\nonumber\\
p_3^{I}(T)&=0
\end{cases}
\end{align}

According to the maximum principle for stochastic differential games (see e.g. [20] , Theorem 2.1 and Theorem 2.3) the problem of finding a Nash equilibrium 
for the two performances $J^{M}(\kappa,\beta), J^{I}(\kappa,\beta)$ can (under some conditions) be reduced to finding a Nash equilibrium for the two Hamiltonians $H^{M},H^{I}$. Thus we proceed to maximize $H^{M}(\kappa,\beta)$ with respect to $\kappa$ for each given $\beta$, and then to maximize $H^{I}(\kappa,\beta)$ with respecty to  $\beta$ for each $\kappa$:
\vskip 0.3cm
For each $t$ the map 
$$\kappa \mapsto H^{M}(t,y_t,m_t,\gamma_t,\kappa,p^{M}(t),q^{M}(t))$$
is convex and therefore it achieves its maximum $\kappa=\hat{\kappa}$ at the boundary, i.e. when
$$\hat{\kappa}=0 \text{ or } \hat{\kappa}= K.$$
Here $K$ is the maximum allowed by the regulator.

The map 
$$\beta \mapsto H^{I}(t,y_t,m_t,\gamma_t,\kappa,\beta,p^{I}(t),q^{I}(t))$$
has a critical point when
\begin{align}
\beta=\hat{\beta}(t)=- \frac{\big[ (v-m_t-\kappa(T-t)y_t)^2+(v-m_t-\kappa(T-t)y_t)p_1^{I}(t)+\frac{\gamma_t}{\sigma_t} q_2(t)\big]\sigma_t^2}{2[\gamma_t(v-m_t)p_2^{I}(t) - \gamma_t^2 p_3(t)]}
\end{align}

We conclude the following:

\begin{thm}
Suppose $(\kappa^*,\beta^*)$ is a Nash equilibrium for Problem 3.1. Then 
$$\kappa^*=0 \text{ or } \kappa^*=K,$$
and the optimal $\beta^*$ is given in feedback form by
\begin{align}
\beta^*(t)= \frac{(v-m_t^*-\kappa^*(T-t)y_t^*)^2+(v-m_t^*-\kappa^*(T-t)y_t^*)(p_1^{I})^*(t)+\frac{\gamma_t^*}{\sigma_t} (q_2^{I})^*(t)}{2[\frac{\gamma_t^2}{\sigma_t^2} (p_3^{I})^*(t) - \frac{\gamma_t^*}{\sigma_t^2}(v-m_t^*)(p_2^{I})^*(t)]}
\end{align}
where $y_t^*, m_t^*, \gamma_t^*, (p_1^{I})^*(t),(p_2^{I})^*(t),(p_3^{I})^*(t),(q_2^{I})^*(t)$ are the system values corresponding to the controls $\kappa^*,\beta^*$.
\end{thm}

\section{Appendix 3.\\
The Bellman approach.}

It may be instructive to see what the dynamic programming approach gives in the present situation. In particular, this may throw some light on the  interpretations of the adjoint variables in Theorem 3.2. In doing so, we take into account our previous remarks made just prior to equation (\ref{E: yt}) in Section 2.2, which tells us to focus on the insider's profit only, since the market maker does not act  strategically, he only trades 'fees'.

Let us for short use the notation $x_t = (y_t, m_t, \gamma_t)$ for the system. The performance functional is given in (3.5), and the maximal profit of the insider is
\begin{equation}\label{E: vf}
J^{I}(x) = w_0^{I} + \text{sup}_{\beta}E\Big[\int_0^T\Big(\tilde{v} - m_s - k_sy_s\big)^2\beta_sds\big].
\end{equation}
With $J^{I}(x,t)$ equal to  the optimal wealth remaining  at time $t$ in state $x$, the Bellman equation can be written 
\begin{equation}\label{E: BE}
\text{sup}_{\beta}\Big\{(v -  m_t - k_ty_t)^2\beta_t + L^{\beta}(J^{I}(x,t))\Big\} = 0,
\end{equation}
where 
\begin{multline}\label{E: L}
L^{\beta}(J^{I}(x,t)) = \frac{\partial J^{I}(x,t)}{\partial t} + (v -  m_t - k_ty_t)\beta_t\frac{\partial J^{I}(x,t)}{\partial y} +\frac{\gamma_t\beta_t}{\sigma_t^2}(v - m_t)\beta_t\frac{\partial J^{I}(x,t)}{\partial m}\\
- \frac{\beta_t^2\gamma_t^2}{\sigma_t^2}\frac{\partial J^{I}(x,t)}{\partial \gamma} + \sigma_t\frac{\partial^2 J^{I}(x,t)}{\partial y^2} + \sigma_t\frac{\partial^2 J^{I}(x,t)}{\partial m^2} + 2\sigma_t\frac{\partial^2 J^{I}(x,t)}{\partial y \partial m}.
\end{multline}
Let us  first address the maximization problem in (\ref{E: BE}). The first order condition  in $\beta_t$ can be written 
\[
(v -  m_t - k_ty_t)^2 + (v -  m_t - k_ty_t)\frac{\partial J^{I}(x,t)}{\partial y} + 2\beta_t\frac{\gamma_t}{\sigma_t^2}(v - m_t)\frac{\partial J^{I}(x,t)}{\partial m}
\]
\[
- 2\beta_t \frac{\gamma_t^2}{\sigma_t^2}\frac{\partial J^{I}(x,t)}{\partial \gamma}  = 0.
\]
This gives the optimal $\beta_t^*$ in terms of the function $J^{I}(x,t)$ (i.e., its partial derivatives) as follows
\begin{equation}\label{E: pobeta}
\beta_t^* = \frac{(v -  m_t - \kappa(T-t)y_t)^2 + (v -  m_t - \kappa(T-t)y_t)\frac{\partial J^{I}(x,t)}{\partial y}}{2[\frac{\gamma_t^2}{\sigma_t^2}\frac{\partial J^{I}(x,t)}{\partial \gamma} - \frac{\gamma_t}{\sigma_t^2}(v - m_t)\frac{\partial J^{I}(x,t)}{\partial m}]}.
\end{equation}
It remains to determine the function $J^{I}(x,t)$. 

Comparing this expression for the optimal trading intensity of the insider with the corresponding  expression in  (3.14) derived using the stochastic maximum principle for stochastic differentiable games, we notice that the adjoint variables $p_i^{I}(t)$, $i=1, 2, 3$ and $q_2^{I}(t)$
can be expressed as follows
\begin{align*}
(p_1^{I})^*(t) &= \frac{\partial J^{I}(x,t)}{\partial y}, &  (p_2^{I})^*(t) &= \frac{\partial J^{I}(x,t)}{\partial m},\\ 
 (p_3^{I})^*(t) &= \frac{\partial J^{I}(x,t)}{\partial \gamma}, & (q_2^{I})^*(t) &= 0.
\end{align*}
These relationships give us more insight into how to interpret these adjoint variables, namely for the  $p_i^{I}(t)$'s as marginal profits with respect to the state variables of the problem at any time $t \in (0,T)$. It also says that the adjoint variable  $(p_2^{I})^*(t)$ has no diffusion term, so this is  a finite variation process.

Notice that it is not normally the case that the adjoint variables in the stochastic maximum principle can be interpreted as 'shadow prices' as we can here. When the state variables have different volatilities, or driven by different Brownian motions, this will in general no longer be true (see e.g., Yong and Zhou (1999)).

The insider's indirect utility function $J^I(x, t) = \int_t^T\beta_s(\gamma_s(\beta) + k_s^2V_s)ds$, can be written as a function of the state $x$ variable and time $t$ as indicated by this notation, and from a conjectured functional form we may attempt to solve the Bellman equation, and  proceed to solution for the trading intensity $\beta$. The result of this  can in its turn be used to address the problem of Appendix 2. However, here we choose to stop and leave this for future research.

\section{Appendix 4.\\
A connection to filtering theory.}

The results of Section 3.1 can alternatively be derived  using filter theory as follows: We first consider the process $y$ for $k = 0$. Then $dy_t = (\tilde{v} - E(\tilde{v}|\mathcal{F}_t^y)\beta_t + \sigma_tdB_t$. From filtering theory (see Allinger and Mitter (1981)) we then know that $y$ generates the same filtration as $\hat{y}$, i.e., $\mathcal{F}_t^{\hat{y}} = \mathcal{F}_t^y $,  and that $\tilde{y}$  defined by $d\tilde{y}_t := \frac{1}{\sigma_t}dy_t:= db_t$ is a Brownian motion   with respect to the  information filtration  $\mathcal{F}_t^y$.  \footnote{The result by Allinger and Mitter proved a long-standing conjecture by Kailath.}

Employing this result to our situation when $k \ne 0$, , we obtain that
\[
\frac{1}{\sigma_t}\{dy_t + k_t\beta_ty_tdt\} := db_t
\]
for an $\mathcal{F}_t^y$-Brownian motion  $b_t$. We may express the total order process $y$ as follows
\[
dy_t = -k_t\beta_ty_tdt + \sigma_tdb_t.
\]
We now employ standard results for Gaussian processes to find $\mu_t:= E(y_t)$ and $V(t):= E(y_t^2)$ for all $t \in [0, T]$. Using Karatzas and Schreeve (1985), we have that  $\mu(t) = E(y_t) = 0$ for all $t$ provided $y_0 = 0$, and the following  first order non-homogeneous ordinary linear differential equation for the variance $V(t) = E(y_t^2)$,
\[
\frac{dV(t)}{dt} = -2k_t\beta_tV(t) + \sigma_t^2,  \quad V(0) = 0
\]
which has  the solution
\begin{equation}\label{E: 6}
V(t) = E(y_t^2) = e^{-2\int_0^tk_s\beta_sds}\int_0^t\sigma_s^2 e^{2\int_0^sk_r\beta_rdr}ds.
\end{equation}
This is (\ref{E: eysquare}).

\paragraph{Acknowledgments}

\end{document}